\documentclass{bioinfo}
\usepackage{url}
\newcommand{\subs}[1]{_{\mathrm{#1}}}

\newcommand{\eq}[1]{\begin{equation}#1\end{equation}}

\newcommand{\eqa}[1]{\begin{eqnarray}#1\end{eqnarray}}
\newcommand{\refeq}[1]{Eq \ref{eq-#1}}
\newcommand{\reftab}[1]{Table \ref{tab-#1}}
\newcommand{\refig}[1]{Figure \ref{fig-#1}}
\newcommand{\refsec}[1]{Section \ref{sec-#1}}
\newcommand{\refssec}[1]{Section \ref{ssec-#1}}

\newcommand{\refm}{Methods}
\newcommand{\cm}[1]{}
\renewcommand{\vec}[1]{\mathbf{#1}}
\newcommand{\textpkg}[1]{\texttt{#1}}
\def\cor{\mathrm{Cor}}
\def\bbR{\mathbb{R}}
\def\bbS{\mathbb{S}}
\def\N{\mathcal{N}}
\def\yres{\vec y\subs{res}}

\newlength{\widthcq}
\newlength{\widthct}
\newcommand{\argmin}{\operatornamewithlimits{argmin}}
\graphicspath{{./figures/}{./}}
\newtheorem{prop}{Proposition}
\newtheorem{remark}{Remark}
\newcommand{\refremark}[1]{Remark \ref{remark-#1}}

\def\pkgname{lassopv}
\def\Pkgname{Lassopv}
\def\atitle{Controlling false discoveries in Bayesian gene networks with lasso regression p-values}
\def\pv{$p$-value}

\def\lp{\textpkg{\pkgname}}
\def\Lp{\textpkg{\Pkgname}}
\def\ct{\textpkg{covTest}}
\def\Ct{\textpkg{CovTest}}
\def\si{\textpkg{selectiveInference}}
\def\Si{\textpkg{SelectiveInference}}
\def\Rt{R${}^2$}
\def\CDF{\mathrm{CDF}}

\copyrightyear{2015} \pubyear{2015}

\access{Advance Access Publication Date: Day Month Year}
\appnotes{Manuscript Category}

\begin{document}

\def\spacingset#1{\renewcommand{\baselinestretch}%
{#1}\small\normalsize} \spacingset{1}
\firstpage{1}

\subtitle{Subject Section}

\title[\atitle]{\atitle}
\author[Wang \textit{et~al}.]{Lingfei Wang\,$^\ast$ and Tom Michoel}
\address{Department of Genetics \& Genomics, The Roslin Institute, The University of Edinburgh, Easter Bush, Midlothian EH25 9RG, UK.}

\corresp{$^\ast$To whom correspondence should be addressed.}

\history{Received on XXXXX; revised on XXXXX; accepted on XXXXX}

\editor{Associate Editor: XXXXXXX}

\abstract{\textbf{Motivation:} Bayesian networks can represent directed gene regulations and therefore are favored over co-expression networks. However, hardly any Bayesian network study concerns the false discovery control (FDC) of network edges, leading to low accuracies due to systematic biases from inconsistent false discovery levels in the same study.\\
\textbf{Results:} We design four empirical tests to examine the FDC of Bayesian networks from three \pv\ based lasso regression variable selections --- two existing and one we originate. Our method, \lp, computes \pv{}s for the critical regularization strength at which a predictor starts to contribute to lasso regression. Using null and Geuvadis datasets, we find that \lp\ obtains optimal FDC in Bayesian gene networks, whilst existing methods have defective \pv{}s. The FDC concept and tests extend to most network inference scenarios and will guide the design and improvement of new and existing methods. Our novel variable selection method with lasso regression also allows FDC on other datasets and questions, even beyond network inference and computational biology.\\
\textbf{Availability:} \Lp\ is implemented in R and freely available at \url{https://github.com/lingfeiwang/lassopv} and \url{https://cran.r-project.org/package=lassopv}.\\
\textbf{Contact:} \href{Lingfei.Wang@roslin.ed.ac.uk}{Lingfei.Wang@roslin.ed.ac.uk}\\
\textbf{Supplementary information:} Supplementary data are available at \textit{Bioinformatics}
online.}

\maketitle

\section{Introduction}
\label{sec-introduction}
The reconstruction of gene regulation networks from gene expression data has been one of the major interests and challenges in computational and systems biology (\cite{schadt_molecular_2009,civelek_systems_2014}). As opposed to co-expression networks, directed networks are specific on the source and target of gene regulations and, consequently, have attracted increasing attention. Among them, Bayesian networks model both the correlation structure among gene expression profiles and the conditional independencies between them, using directed acyclic graphs (DAGs), such as in \cite{friedman_learning_1999,friedman_using_2000,zhu_integrating_2008,zhang_integrated_2013}. An accurate Bayesian network can significantly improve downstream analyses and experimental validations by predicting the master regulators and the genes responding to them in specific experimental conditions \cite{zhu_integrating_2008,zhang_integrated_2013,talukdar2016}

Despite such advantages of Bayesian networks over co-expression networks, its false discovery control (FDC), i.e.\ setting a network sparsity threshold such that the expected number of false positive edges is controlled, has been left mostly untouched as a statistical hypothesis testing question. Co-expression networks contain a \pv{} for every pair of genes, from which a threshold for FDC can be chosen (e.g.\ \cite{langfelder_wgcna:_2008}), despite the low specificity. However, no Bayesian network inference method to our knowledge provides a spectrum of continuous values (like \pv) for FDC. Instead, Bayesian network inference has been regarded as either a mathematical optimization problem or a multi-step, multi-parametric test \cite{kalisch_estimating_2007,scutari_learning_2010}, so FDC has become very difficult. The importance of FDC, or even FDC itself, has been mostly overlooked in Bayesian networks.

The consequences of lacking FDC are obvious. Without a \pv-like variable to control the false discovery rate, any choice of network sparsity is hard to justify statistically. An even more detrimental consequence is the sub-optimal inference accuracy due to one group of interactions being favored over the rest by statistical bias, as discussed in detail in \refsec{approach}. Testing and achieving FDC would allow us to evaluate and improve network inference methods.

To control the false discovery, in this paper we consider the Bayesian network inference when a natural ordering of genes is given. In this case, Bayesian network inference reduces to a series of individual variable selection problems \cite{koller_probabilistic_2009}. Unlike traditional likelihood optimization or sampling algorithms, these approaches also easily scale to large systems involving thousands of genes. Furthermore, they are not hampered by the fact that DAGs form equivalence classes, in which individual DAGs cannot be distinguished by observational data alone \cite{chickering_learning_2002}. Natural gene orderings or (dense) prior DAGs are typically obtained from external information (e.g.\ signalling pathways or known transcription factor binding target information)\cite{Shojaie:2010}, or, in a systems genetics context, can be inferred using expression quantitative trait loci in the neighborhood of each gene's transcription start site (cis-eQTL) as causal anchors \cite{chen_harnessing_2007,millstein_disentangling_2009,wang_efficient_2017}. By then applying regression methods, particularly the lasso \cite{Tibshirani:1996,efron_least_2004,meinshausen2006high} which favors sparse solutions, we can perform variable selection on the candidate regulators of each gene, and obtain sparse, genome-scale, and high-quality Bayesian networks \cite{schmidt:2007,Shojaie:2010,aragam2014concave}.

There have been several attempts to repurpose lasso regression for statistical variable selection. Cross-validation, the standard approach in predictive lasso regression, is computationally intensive and disregards variable selection FDC \cite{Shojaie:2010}. Scaling regularization strengths with the number of candidate parental genes, as proposed in \cite{schmidt:2007}, offsets more parents with stronger selection, and can provide an upper bound for false discovery. Without a lower bound, however, its over-conservative FDC is still subject to biases dependent on the number of parental genes. None of these methods could demonstrate consistent FDC in network inference or variable selection.

Recent developments in sequential hypothesis testing regarded lasso regression as hypothesis testing in variable selection \cite{Lockhart:2014,Taylor:2016,Lee:2016}. By considering every candidate predictor separately, the null hypothesis assumes an independent predictor, from which the \pv\ of specific lasso statistics of interest may be obtained. Accurate \pv{}s of lasso regression variable selection would make possible FDC in Bayesian networks, like in co-expression networks and association studies. Consistent FDCs across all target genes would allow for optimal inference accuracy and a self-justified network sparsity.

In this paper, we develop the software \lp\ to compute the \pv{} of the critical regularization strength at which every predictor starts to contribute to lasso regression for the first time \cite{wang_lassopv_2017}. This choice of statistic is based on the established criterion that predictors with nonzero contributions are significant, and more so at stronger regularizations. We also propose four statistical tests to empirically evaluate the FDC of a reconstructed Bayesian network. To compare \lp's FDC against existing methods (\ct\ \cite{Lockhart:2014} and \si\ \cite{Taylor:2016}), we apply these statistical tests on their reconstructed networks from four low-dimensional and high-dimensional, simulated null and real biological datasets \cite{Lappalainen:2013}. We also demonstrate Bayesian networks' advantage over co-expression networks in avoiding indirect regulations and confounding. \Lp\ is publicly available in R at \url{https://github.com/lingfeiwang/lassopv} and \url{https://cran.r-project.org/package=lassopv}.


\section{Approach}
\label{sec-approach}
The critical question in Bayesian network inference is orienting the regulations, which can be achieved with three broad categories of methods. The first, known as the PC algorithm \cite{ramsey_adjacency-faithfulness_2006,kalisch_estimating_2007}, tries to identify the V-shaped interactions $G_1\rightarrow G_2\leftarrow G_3$ between three genes $G_1,G_2,G_3$, and then propagates the orientations on the co-expression network. The second method treats every Bayesian network as a regression model, and seeks the best predictive network(s) with Monte Carlo simulations \cite{scutari_learning_2010}. The third method, as discussed before, introduces external information to order all genes, and then network inference becomes a series of regression problems of every gene on its predecessors in the ordering. Hybrid methods have also been developed.

Regardless of the category, however, the form of their products is identical --- a list of gene regulations that comprise the Bayesian network at any given significance level, regularization strength, or other custom cutoff value. This threshold determines network sparsity: starting with no interaction, and as the threshold changes monotonically, the network gradually includes more and more interactions, and finally reaches a full network (with $n(n-1)/2$ interactions for $n$ genes). Therefore, every Bayesian network inference method, regardless of the category, consists effectively of two separate and sequential questions --- first how to orient all regulations, answered by the full network, and then what value to assign to each regulation to rank their significance, answered by the sequence and/or critical threshold at which they become significant in the network.

In this paper, we focus on the FDC of the latter question of value assignment. To understand how the lack of FDC can undermine inference accuracy, as mentioned in \refsec{introduction}, consider an example association study for multiple diseases whose sample sizes differ. If the correlation coefficient or odds ratio was used as a significance measure, the study could not account for the sample size differences, and therefore would bias towards associations with diseases with fewer samples. Such biases lead to inconsistent FDC levels between different groups of tests (here diseases), and therefore reduce the overall accuracy. The proper significance measure in the example would be the \pv{} (or false discovery rate) for observing a correlation coefficient or odds ratio value, which can ensure a unified FDC and an optimal accuracy across multiple tests. In network inference, statistical tests are composite, sequential, and much more complicated than simple pairwise correlations. Therefore, such biases cannot be assumed absent with merely equal sample sizes, but should be tested.

Testing FDC bias in network inference is equivalent to testing (pseudo-)random number generators in cryptography. Optimal FDC prevents any bias, and therefore all the non-existent candidate interactions (false cases) would appear identical to the network inference method, which can only assign values that rank them randomly and featurelessly. Although the possible feature space has infinite dimensions, there are typical bias features which network inference methods tend to produce, and therefore should be tested, as for random number generators \cite{bassham_statistical_2010}. For example, after orienting the interactions, the different numbers of candidate incoming interactions for different genes may introduce a bias, especially for regression based network inference.

However, most Bayesian network inferences only produce a list of significant interactions without specifying their values or rankings, posing a major difficulty to the empirical evaluations of FDC. Under the null hypothesis, each interaction has an equal probability of being significant, at any network sparsity. Therefore, for any target gene, the expected number of its significant incoming interactions (i.e.\ false positives) should be the product of that of its potential ones (i.e.\ false cases) and the false positive rate (FPR). As a result, given only a list of significant interactions, a consistent FPR would yield a testable linear relation between the numbers of significant and possible incoming interactions.

Besides the linearity test, we devise additional tests for Bayesian networks of continuous values or rankings of potential interactions, such as those obtained from lasso regression \pv{}s. We also account for other issues in FDC on real data, such as the high dimensionality in genomic data and the ``contamination'' from unknown genuine interactions.

\begin{methods}
\section{Methods}
\subsection{Data\label{ssec-data}}
We used the Geuvadis consortium's gene expression levels and SNPs from lymphoblastoid cell lines of 360 European individuals \cite{Lappalainen:2013} for gene network reconstruction. We limited our analyses to the 3172 genes that possess at least one significant cis-eQTL \cite{Lappalainen:2013}. The expression levels of every gene were converted into following the standard normal distribution by relative ranking \cite{chen_harnessing_2007} prior to subsequent analyses.

We used the most significant cis-eQTL for every gene as the causal anchor \cite{schadt2005integrative,wang_efficient_2017} to infer the probability of directed regulation between all pairs of the 3172 genes, using the function \textpkg{pij\_gassist} in \textpkg{Findr} 1.0.5 \cite{wang_efficient_2017,wang_findr_2017}. Based on the inferred probability of pairwise regulation, we then constructed a greedy, maximal DAG by adding one directed regulation at a time in descending probability order using \textpkg{netr\_one\_greedy} also in \textpkg{Findr} 1.0.5. Edges that were to create a loop were discarded. The maximal DAG of 5,029,206 edges represented the natural prior ordering of the 3172 genes which, together with their expression levels across 360 individuals, formed the input of lasso-based Bayesian network learning.

Based on the above full, high-dimensional Geuvadis dataset, we additionally derived three datasets to evaluate lasso \pv{}s for Bayesian network learning under different conditions:
\begin{itemize}
\item The high-dimensional null dataset of the same dimension (3172 genes from 360 individuals) was constructed by replacing every expression level with a random sample from independent standard normal distribution. This dataset examines the performance of variable selection methods in a high-dimensional null setting.
\item The low-dimensional Geuvadis dataset of the top 150 genes in the prior order from all 360 individuals validates the low-dimensional real performance.
\item The low-dimensional null dataset was similarly constructed from the low-dimensional Geuvadis dataset, by replacing every expression level with a random sample of independent standard normal distribution, to reflect the low-dimensional null scenario.
\end{itemize}

\subsection{Lasso \pv}
Without loss of generality, assume we have $n$ observations of the target variable as $\vec y\in\bbR^n$ and $k$ predictor variables as $\vec X\equiv(\vec x_1,\vec x_2,\dots,\vec x_k)\in\bbR^{n\times k}$, subject to normalization:
\eq{\mathrm{mean}(\vec y)=0,\hspace{3em}\mathrm{and\ }\forall i=1,\dots,k,\hspace{1em}\mathrm{mean}(\vec x_i)=0.}
Parameterized by the regularization strength $\lambda$, the lasso regularization \cite{Tibshirani:1996} solves the optimization problem
\eq{\hat\beta(\lambda)\equiv\argmin_{\beta\in\bbR^k}\frac{1}{2n}||\vec y-\vec X\beta||_2^2+\lambda||\beta||_1,}
and predicts $\vec y$ with the estimator $\hat{\vec y}(\lambda)$ as
\eq{\hat{\vec y}(\lambda)\equiv\vec X\hat\beta(\lambda).}

To compute the \pv\ for predictor $\vec x_i$, first denote its real variance with $\sigma_i^2\equiv\frac1n||\vec x_i||_2^2$. Its null hypothesis $H_0^{(i)}$ can be defined as a uniformly distributed vector on the sphere $\bbS^{n-1}$ with the same variance/radius, \textit{i.e.}
\eq{P\left(\vec x_i=\vec v\mid H_0^{(i)}\right)=\left\{\begin{array}{ll}
\mathrm{const},&||\vec v||_2=\sigma_i\sqrt n,\\
0,&\mathrm{else}.\end{array}\right.\label{eq-h0-s}}
We chose the test statistic as the critical regularization strength $\lambda=\lambda_i$ at which predictor $\vec x_i$ first contributes to predicting $\vec y$, as
\eq{\lambda_i\equiv\sup\{\lambda:\hat\beta_i(\lambda)\ne0\}.}
The one-sided \pv\ of the critical regularization strength $\tilde\lambda_i$ of $\vec x_i$ in the regularization path is the probability of $\lambda_i\ge\tilde\lambda_i$ under the null hypothesis, which can be analytically computed under approximation (\refsec{pv}) as
\eqa{&&P\left(\lambda_i\ge\tilde\lambda_i\mid H_0^{(i)}\right)\nonumber\\
&\approx&2-2\CDF_{t(n-2)}\left(\frac{\tilde\lambda_i}{\sigma_i\sigma_{\yres}(\tilde\lambda_i)}\sqrt{\frac{n-2}{1-\frac{\tilde\lambda_i^2}{\sigma_i^2\sigma_{\yres}^2(\tilde\lambda_i)}}}\right),\label{eq-pv-s}}
where $\CDF_{t(n-2)}$ represents the cumulative distribution function for the Student's $t$-distribution with $n-2$ degrees of freedom, and
\eq{\sigma_{\yres}^2(\lambda)\equiv\frac{1}{n}||\vec y-\hat{\vec y}(\lambda)||_2^2}
is the residue variance of $\vec y$ at any given $\lambda$.

Besides the `\textit{spherical}' null hypothesis above, in \refsec{pv} we also considered the `\textit{normal}' null hypothesis (\refeq{spv-n}) of independent normal distributions with variance $\sigma_i^2$. We developed the package \lp\ \cite{wang_lassopv_2017} to compute the \pv{}s for these null hypotheses.

\subsection{False discovery control using \pv{}s in network inference\label{ssec-fdr}}
To prune a prior DAG of possible regulations as set $\N_0\equiv\{i\rightarrow j\}$ of $n$ genes $i,j\in\{1,\dots,n\}$, we need to compute the \pv\ of every regulation $i\rightarrow j\in \N_0$ as $p_{ij}$, separately for each target node $j$ as a variable selection. Noting that the null \pv{}s follow the standard uniform distribution $p_{ij}\sim U(0,1)$, we evaluated the quality of null \pv{}s and their FDC with the following tests:
\begin{itemize}
\item\textbf{Histogram test}: To evaluate the null \pv{}s as a whole, we visualized the overall histogram of $p_{ij}$ against the standard uniform histogram of the same observation size.
\item\textbf{The Kolmogorov-Smirnov (KS) test}: To evaluate the null \pv{}s separately for each variable selection, we performed the KS test \cite{Kolmogorov:1933,Smirnov:1948} on $p_{ij}$ against the standard uniform distribution (separately for each target gene $j$ but over all its possible regulators $i$). A Manhattan plot then shows the KS test \pv{}s as a function of the number of possible regulators $N_j\equiv\#_i(i\rightarrow j\in \N_0)$. The KS \pv{}s were then compared against the 0.05 significance threshold with Bonferroni correction.
\item\textbf{Linearity test}: To test whether \pv{}s can be compared across different variable selections in network inference, we can choose a significance threshold $p\subs{thres}$ for \pv. Under the null hypothesis, the number of significant regulators for every target gene $i$ would approximately follow the normal distribution $N(N_ip\subs{thres},N_ip\subs{thres}(1-p\subs{thres}))$ (or strictly speaking the binomial distribution $B(N_i,p\subs{thres})$). For a given threshold $p\subs{thres}$, we can visually examine the linear relation between the numbers of possible and significant regulators for each target in a scatter plot.
\item\textbf{\Rt\ test}: Based on the linearity test, we further computed the goodness of linear fit $R^2$ at different significance thresholds, and as a function of the total proportion of significant hypotheses. A higher curve or a larger area under the $R^2$ curve (AUR2) then indicates better FDC. Since the accuracy of small \pv{}s are most important for distinguishing null and non-null cases, partial AUR2s at small \pv{}s were also computed.
\end{itemize}

Although these tests were designed primarily for \pv{}s, the linearity and \Rt\ tests also apply to other continuous statistics straightaway. For the histogram and KS tests, any other continuous statistic (or biased \pv{}s) can also be compared against its (empirical) null distribution, if available. Given only a list of significant edges, the linearity and \Rt\ tests are also test the FDC of the reconstructed binary Bayesian networks.

These tests also assume a null dataset, i.e.\ the absence of any genuine interactions in the network. To extend them to real datasets where sparse real interactions lead to the enrichment of small \pv{}s or other statistics, a proper exclusion cutoff can remove those ``contaminations''. The specific treatments are explained in the relevant results section.

\subsection{Evaluation of existing and new lasso \pv{}s}
We limited our evaluation and comparison among the three R packages that compute lasso \pv{}s:
\begin{itemize}
\item\lp\ 0.2.0: this paper and \cite{wang_lassopv_2018},
\item\ct\ 1.02: covariance test on the local loss of explained variance when a predictor is removed \cite{Lockhart:2014}, and
\item\si\ 1.2.0: post-selection inference based on sequential hypothesis testing (at multiple values of its parameter $\lambda$, \cite{Taylor:2016,Lee:2016}).
\end{itemize}
Methods that cannot give \pv{}s were not included, such as \cite{Shojaie:2010} (\textpkg{knockoff}) and \cite{Barber:2015}.

We evaluated each method on each dataset. Given the inferred maximal DAG, we performed one variable selection per target gene, by computing the lasso \pv{}s of all its potential regulators. Using tests in \refssec{fdr}, we then evaluated the FDC of \pv{}s from different packages and different datasets.

\end{methods}

\section{Results\label{sec-results}}
\subsection{Lassopv obtained uniformly distributed \pv{}s on the low-dimensional null dataset}
The signalling pathways or gene regulatory networks of biological systems can be modeled as a sparse DAG or Bayesian network, inferred from observational gene expression data \cite{friedman2004inferring}. When a superset DAG or a natural ordering of (gene) nodes is given, the problem reduces to a series of variable selections, where each node is regressed on its parents (i.e.\ potential regulators) in the superset DAG \cite{Shojaie:2010,aragam2014concave}. An accurate \pv{} based FDC would allow for a uniform significance threshold \emph{across} the regressions, and therefore an optimal overall accuracy for network inference.

\begin{figure*}[!tpb]
\center
\begin{tabular}{lcccc}
&\lp&\ct&\multicolumn{2}{c}{\si}\\
&&&$\lambda=0.001$&$\lambda=0.02$\\
\rotatebox{90}{\hspace{4em}Histogram}&
\includegraphics[width=\widthcq]{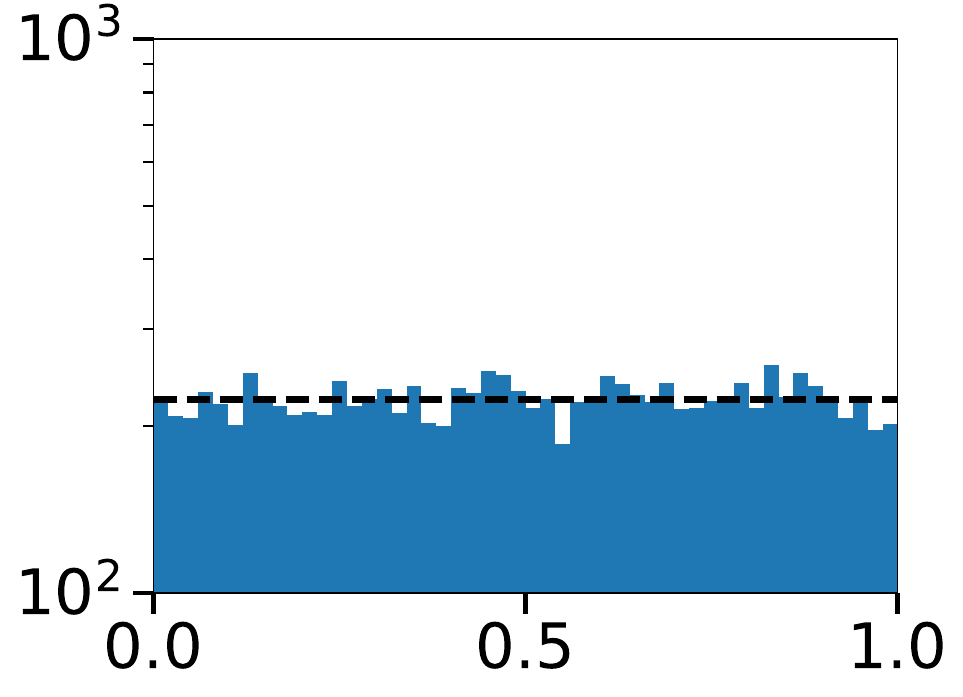}&
\includegraphics[width=\widthcq]{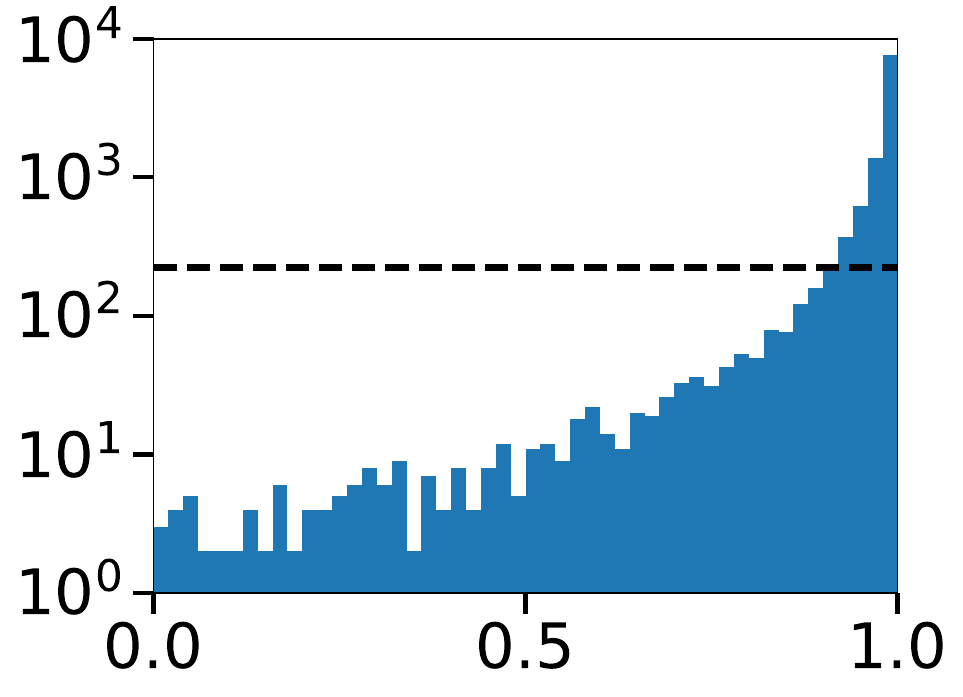}&
\includegraphics[width=\widthcq]{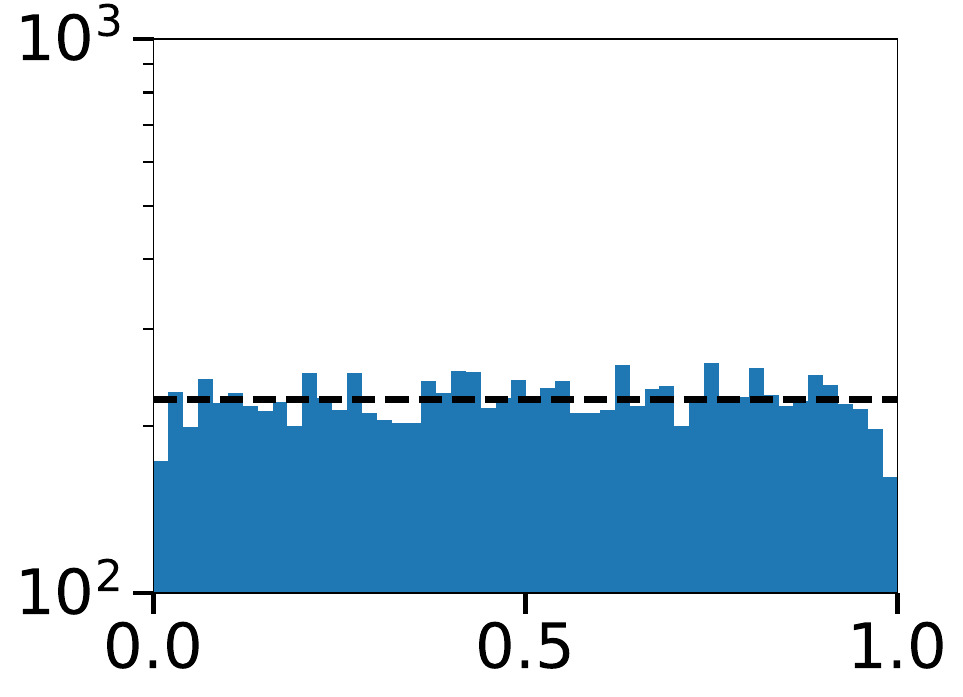}&
\includegraphics[width=\widthcq]{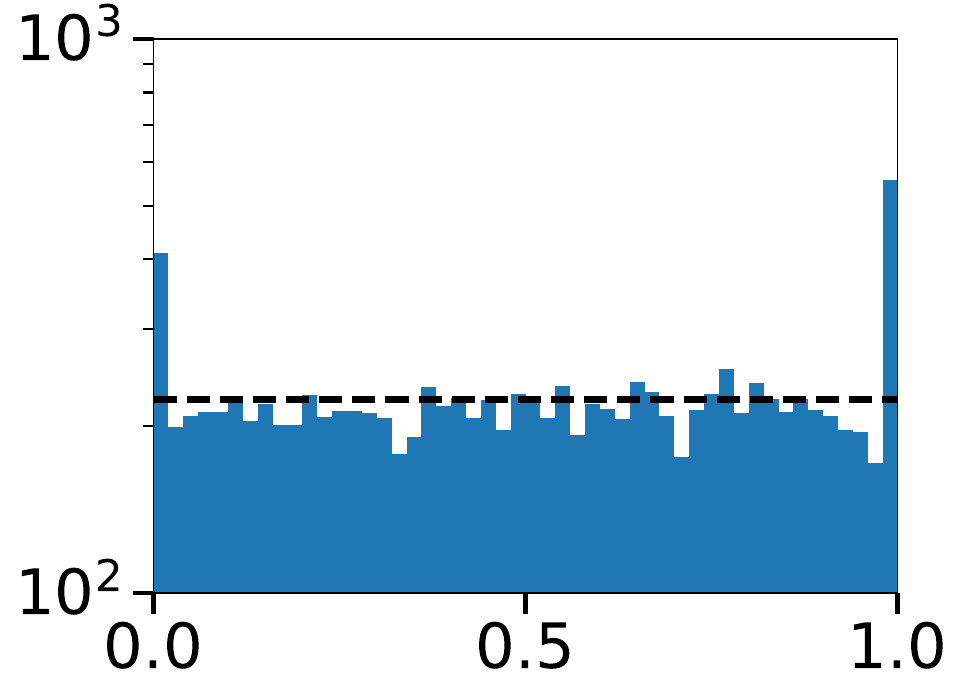}\\
\rotatebox{90}{\hspace{6em}KS}&
\includegraphics[width=\widthcq]{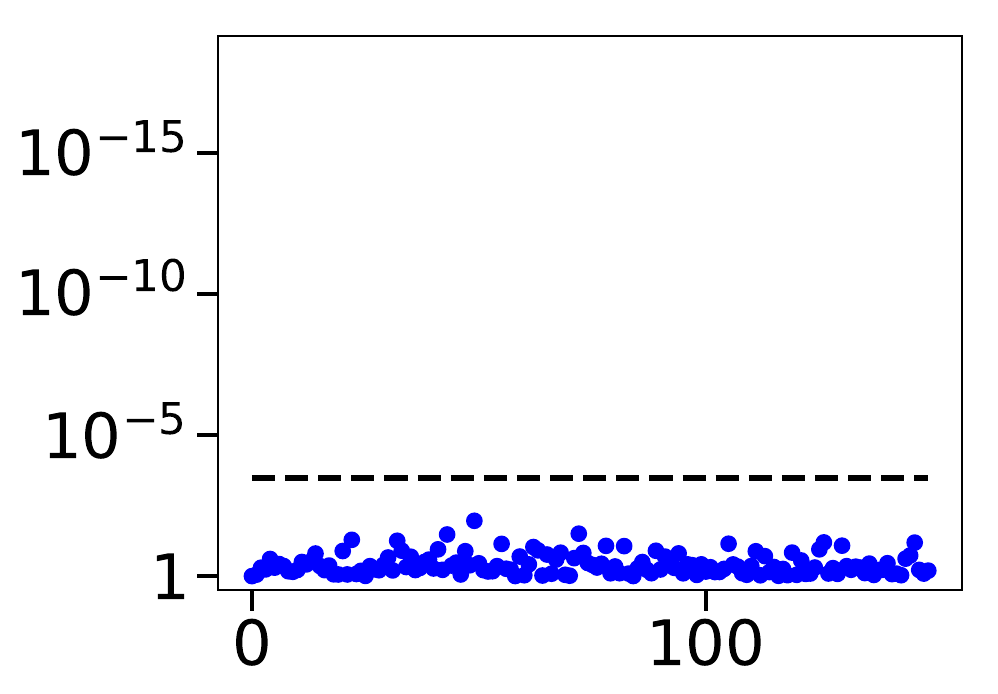}&
\includegraphics[width=\widthcq]{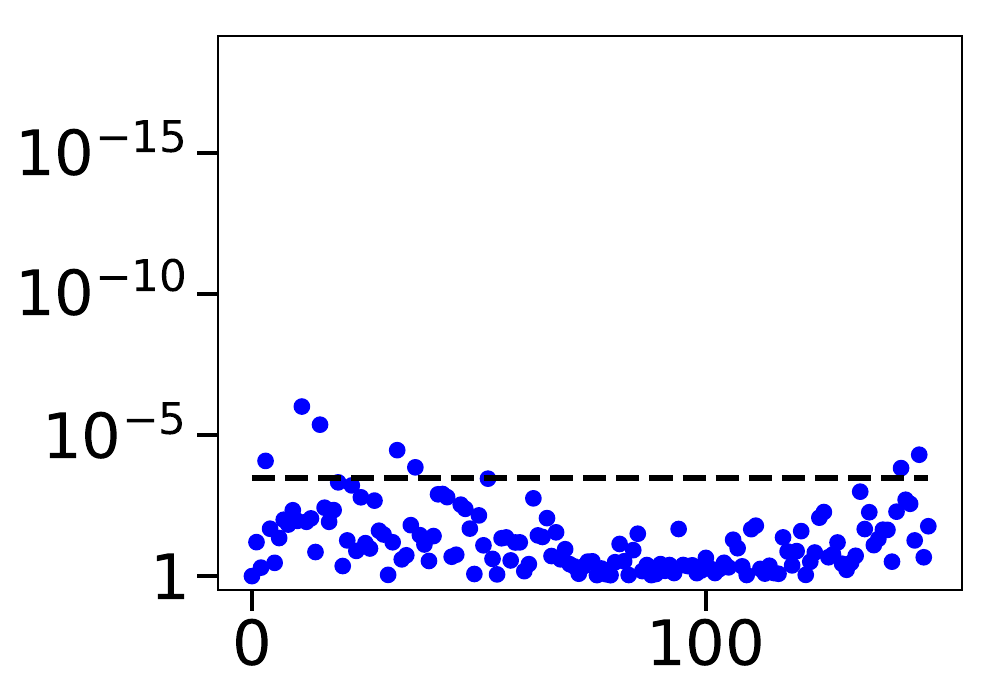}&
\includegraphics[width=\widthcq]{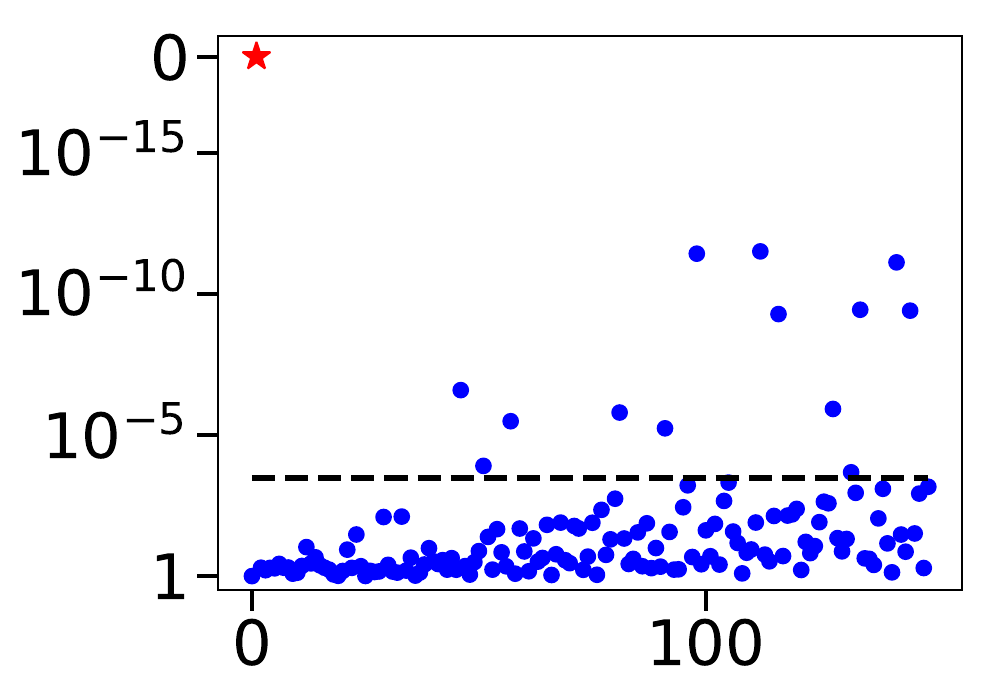}&
\includegraphics[width=\widthcq]{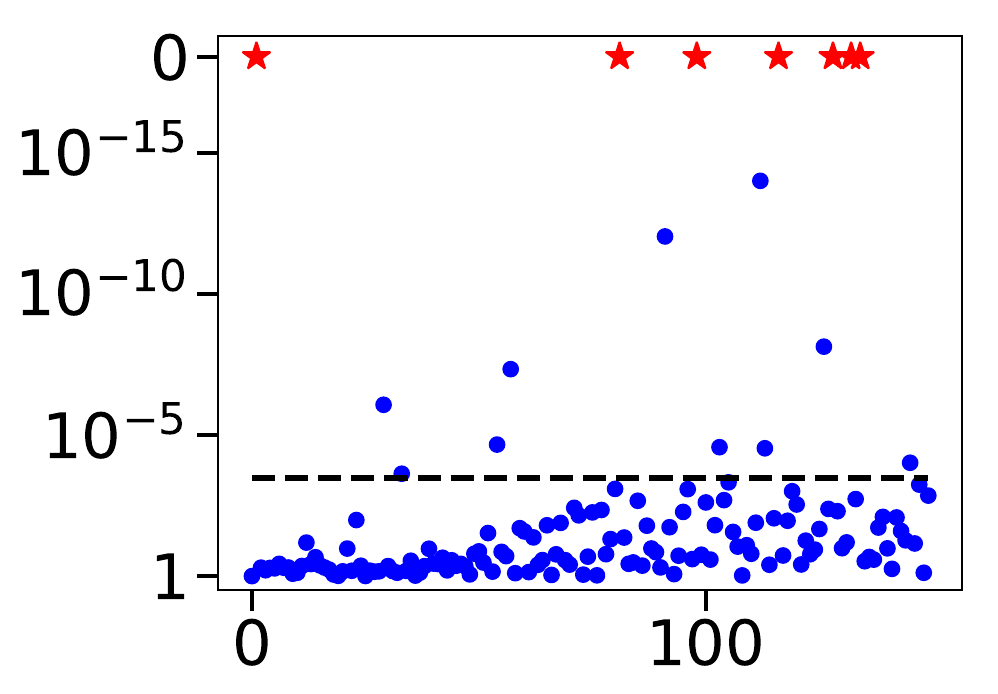}\\
\rotatebox{90}{\hspace{6em}\Rt}&
\includegraphics[width=\widthcq]{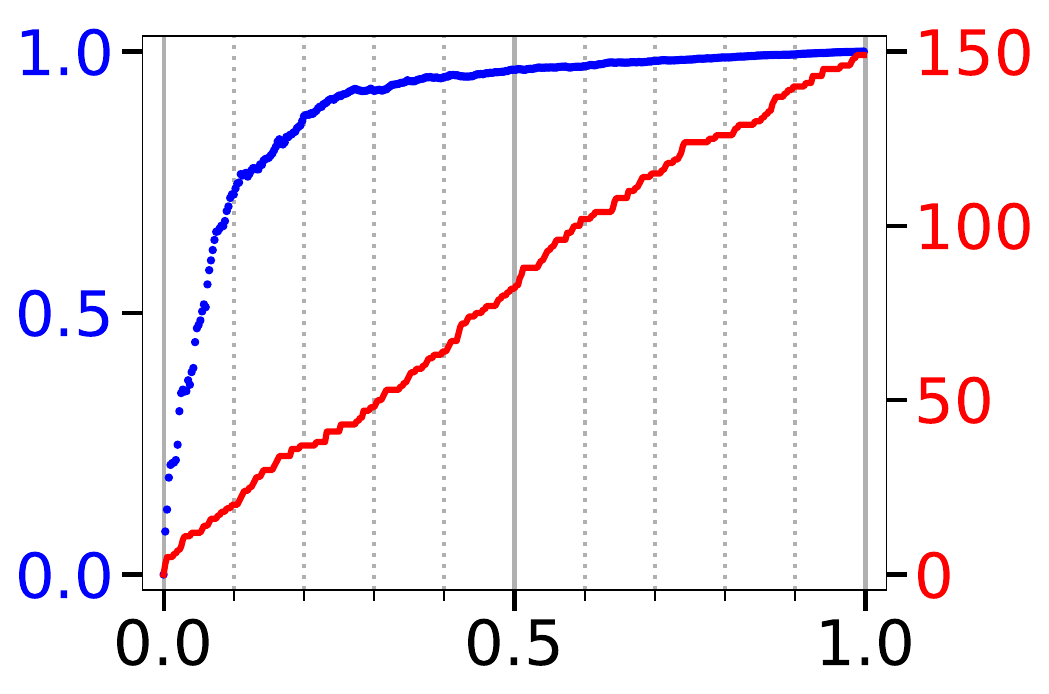}&
\includegraphics[width=\widthcq]{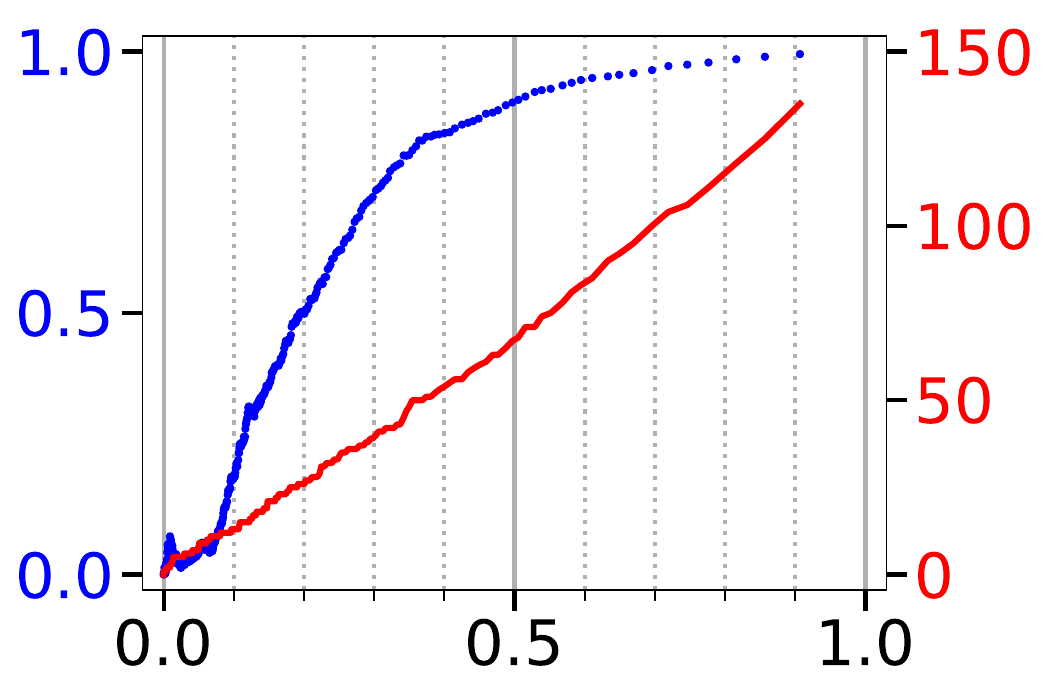}&
\includegraphics[width=\widthcq]{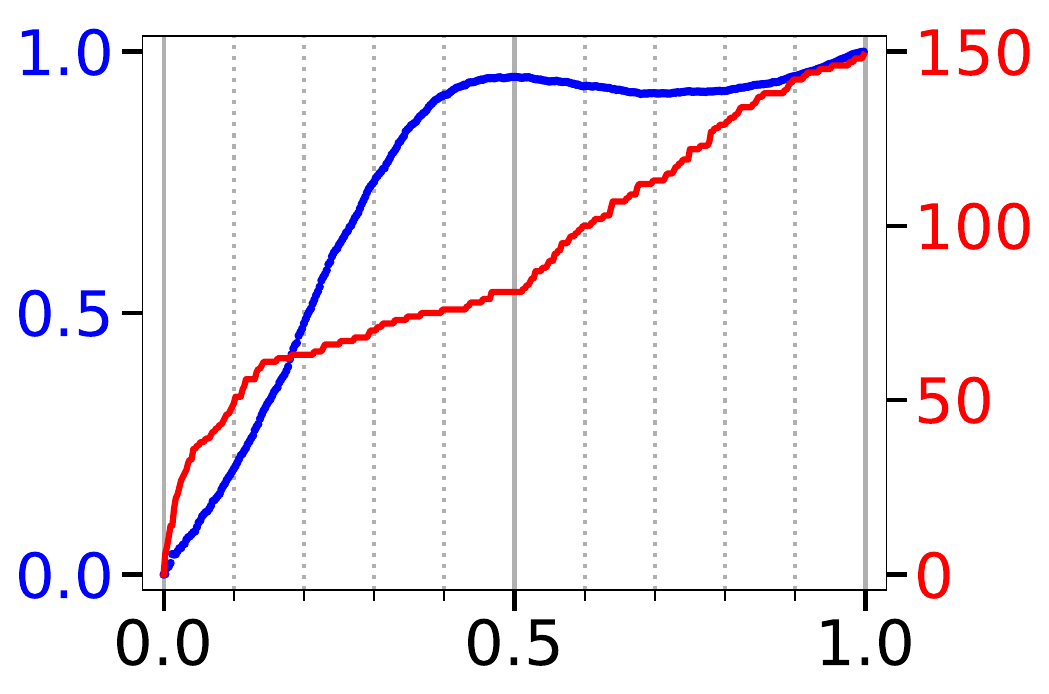}&
\includegraphics[width=\widthcq]{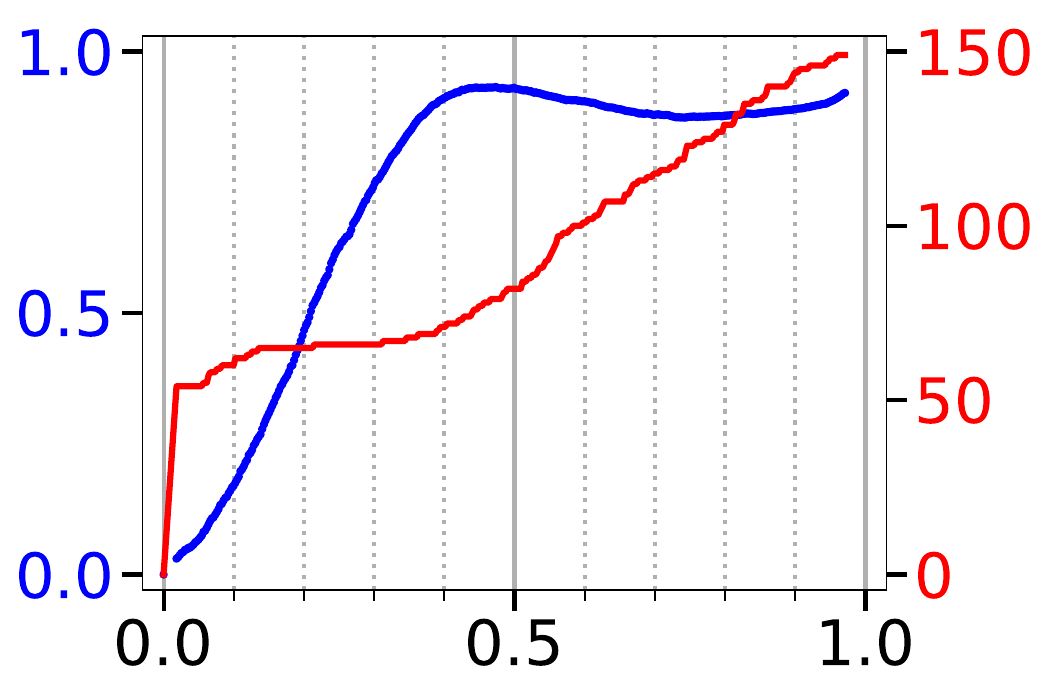}\\
\end{tabular}
\caption{\Lp\ provided accurate FDC than \ct\ and \si\ in the histogram, KS, and \Rt\ tests on low-dimensional null dataset. \textbf{Histogram}: the histograms of lasso \pv{}s, with dashed lines as the perfectly uniform histogram of the same size. \textbf{KS}: the Manhattan plots of KS test \pv{}s ($y$) of lasso \pv{}s against the standard uniform distribution (or the overall empirical distribution for \ct) as a function of the numbers of potential regulators of each gene ($x$). Dashed lines represent the KS test \pv{} 0.05 with Bonferroni correction. High dots and stars (too small for machine precision) indicate \pv{}s significantly deviating from the desired distribution. \textbf{\Rt}: the $R^2$ goodness of fit (blue, left $y$) and the maximum number of significant regulators for all genes (red, right $y$) as functions of the total proportion of significant regulations ($x$) at different significance thresholds. Higher blue and linear red curves indicate better FDC across the variable selection tasks. See \refm{} for test methods.\label{fig-reducednull}}
\end{figure*}

To evaluate the FDC of different lasso \pv\ methods in network inference, we first looked into the baby problem with low-dimensional null datasets and used \lp, \ct, and \si\ to prune the maximal DAG into sparse DAGs. Considering all variable selections together in the histogram test in \refig{reducednull} (histogram), \Lp\ produced uniformly distributed \pv{}s on the null dataset. \Ct\ was highly over-conservative, with over-abundance of \pv{}s towards one. \Si\ had \pv{}s that were over-abundantly small and large at high $\lambda$ but were uniformly distributed at low $\lambda$, in agreement with Figures 3 and 5 in \cite{Taylor:2016}.

We then evaluated variable selections separately for each target gene using the KS test, and \lp\ maintained its optimal FDC (\refig{reducednull}, KS). \Lp\ \pv{}s were consistent with the standard uniform distribution on the low-dimensional null dataset. Although \si\ obtained an overall standard uniform distribution in the histogram test, its null \pv{} distributions were revealed highly non-uniform for many target genes. For \ct, we accounted for its biased \pv{}s by performing the KS test against its overall empirical \pv\ distribution. However, its null \pv\ distributions still differ across different target genes, suggesting that the amount of bias is dependent on the number of possible regulators.

\begin{table}
\center
\caption{\Lp\ outperformed \ct\ and \si\ on AUR2 (\refig{reducednull}) of the low-dimensional null dataset. Partial AUR2s were computed for the given $x$ bounds, and normalized to unit AUR2 at constant function $R^2=1$. Best performers are shown in bold.\label{tab-1}}\vspace{1em}
\begin{tabular}{c|c|c|cc}
&\lp&\ct&\multicolumn{2}{c}{\si}\\
Bound&&&$\lambda=0.001$&0.02\\
\hline
$[0,0.01]$&\textbf{0.12}&0.032&0.010&0.008\\
$[0,0.05]$&\textbf{0.29}&0.031&0.051&0.037\\
$[0,0.2]$&\textbf{0.63}&0.21&0.22&0.19\\
$[0,1]$&\textbf{0.90}&0.76&0.75&0.72
\end{tabular}
\end{table}

To specifically evaluate the FDC in network inference, we then performed the \Rt\ test on the low-dimensional null dataset, which reaffirmed our existing conclusions (\refig{reducednull}, \Rt). \Lp\ obtained a highly linear relation between the numbers of candidate and significant regulators, especially at small \pv{}s, and as opposed to \ct\ and \si. \Si\ assigned too many highly significant regulators to a small number of target genes, which could be the cause of breakdown in the KS test. Their performances were also summarized in full and partial AUR2s in \reftab{1}.

\subsection{Lassopv obtained optimal FDC on the low-dimensional Geuvadis dataset}
To evaluate the FDC of different lasso \pv{} methods on the inference of a real gene regulation network, we first looked into the low-dimensional Geuvadis dataset (\cite{Lappalainen:2013}, \refssec{data}). Despite our lack of knowledge on the groundtruth of the gene regulation network, the sparse nature of incoming regulations for every gene still makes the FDC evaluation possible using the same tests we applied on the null dataset.

\begin{figure*}[!tpb]
\center
\begin{tabular}{lcccc}
&\lp&\ct&\multicolumn{2}{c}{\si}\\
&&&$\lambda=0.001$&$\lambda=0.02$\\
\rotatebox{90}{\hspace{4em}Histogram}&
\includegraphics[width=\widthcq]{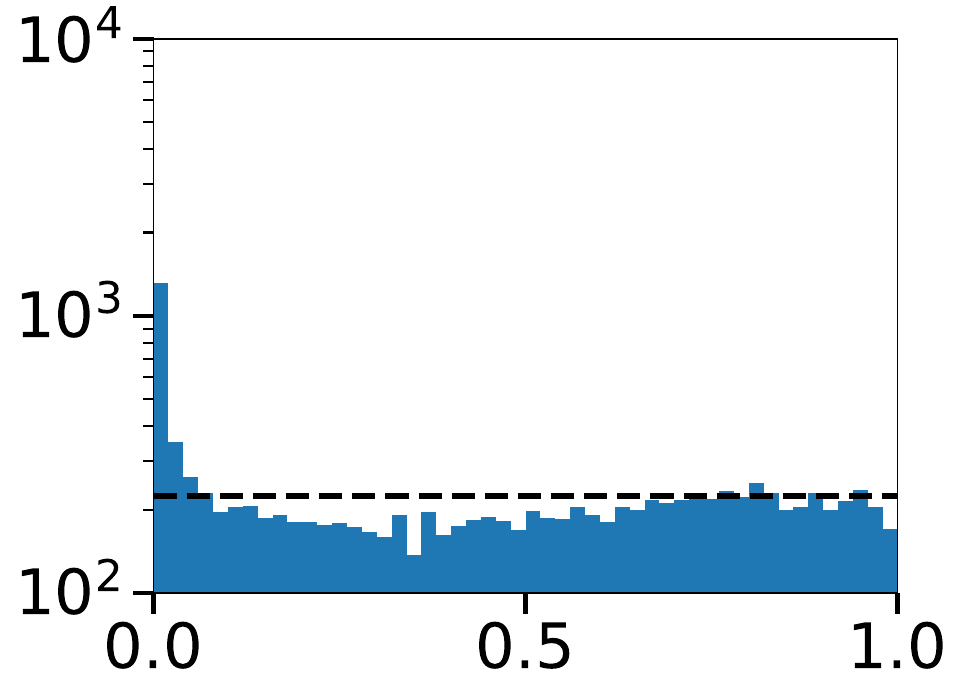}&
\includegraphics[width=\widthcq]{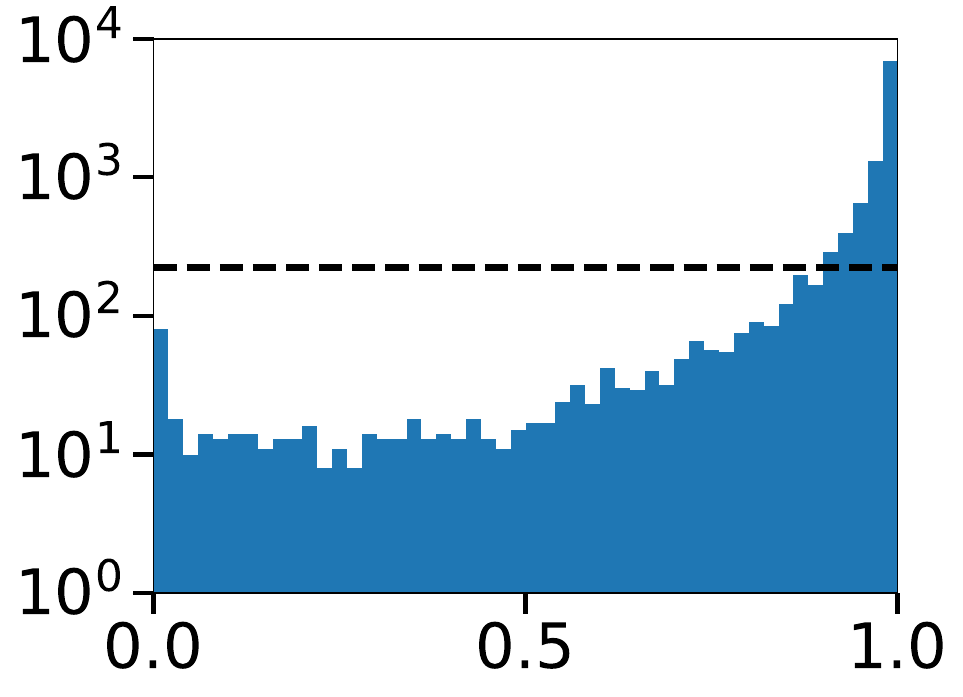}&
\includegraphics[width=\widthcq]{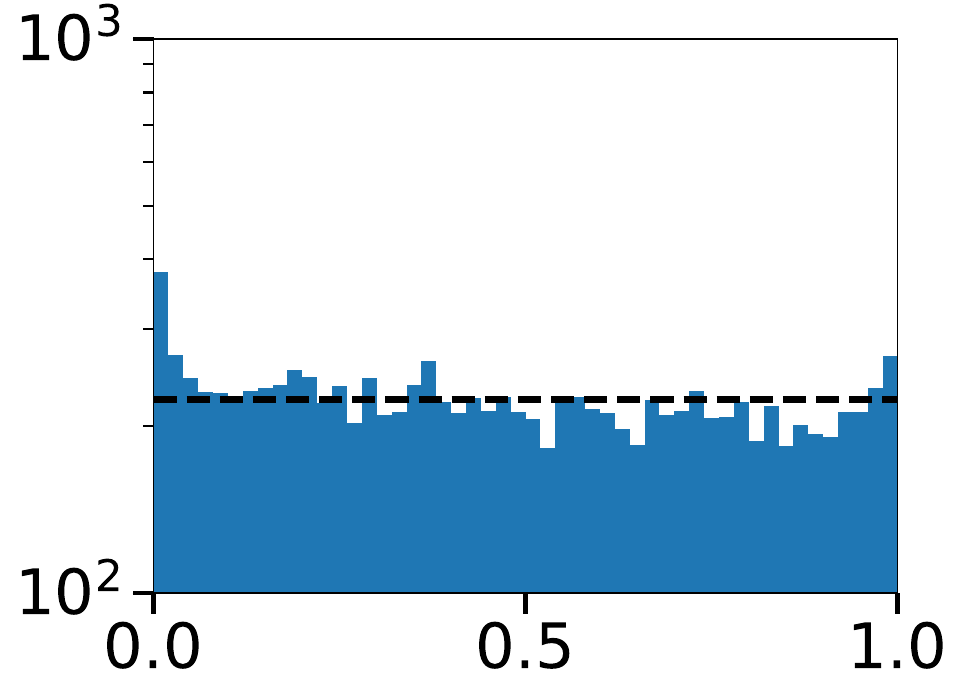}&
\includegraphics[width=\widthcq]{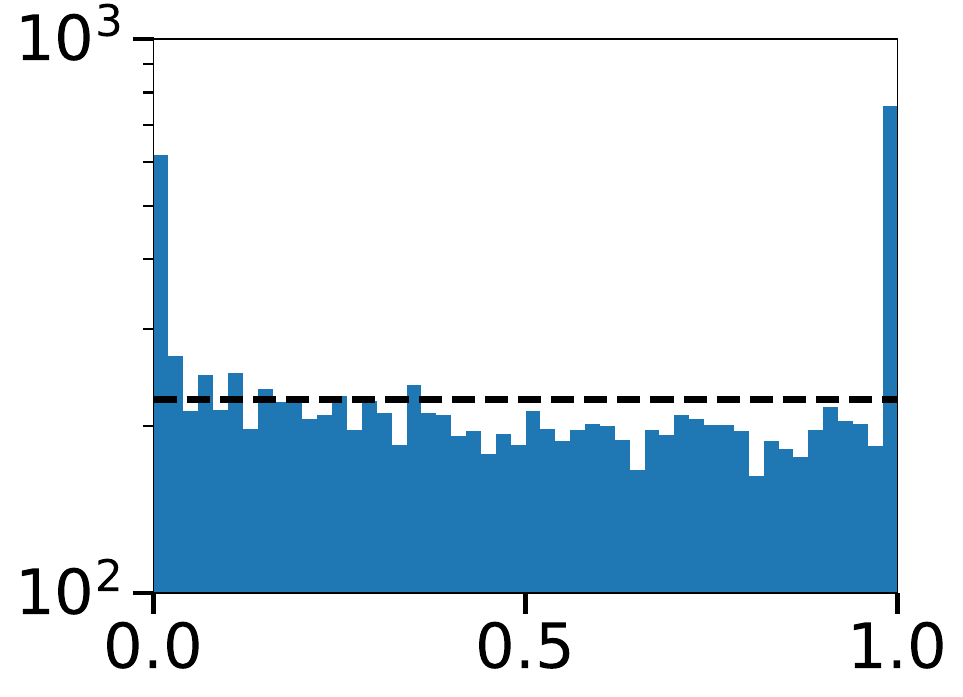}\\
\rotatebox{90}{\hspace{6em}KS}&
\includegraphics[width=\widthcq]{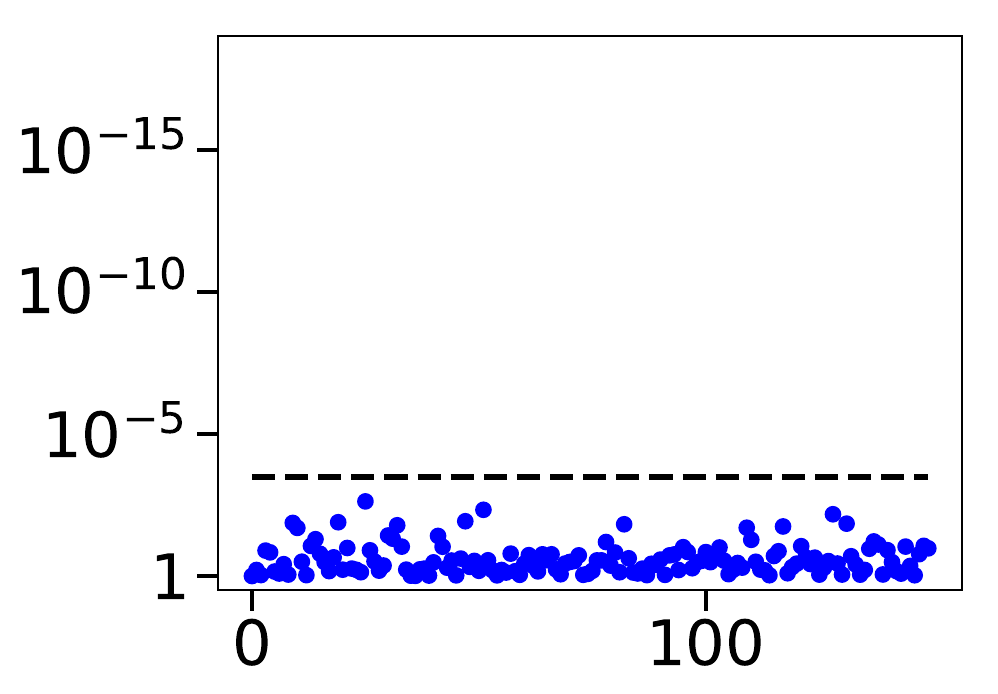}&
\includegraphics[width=\widthcq]{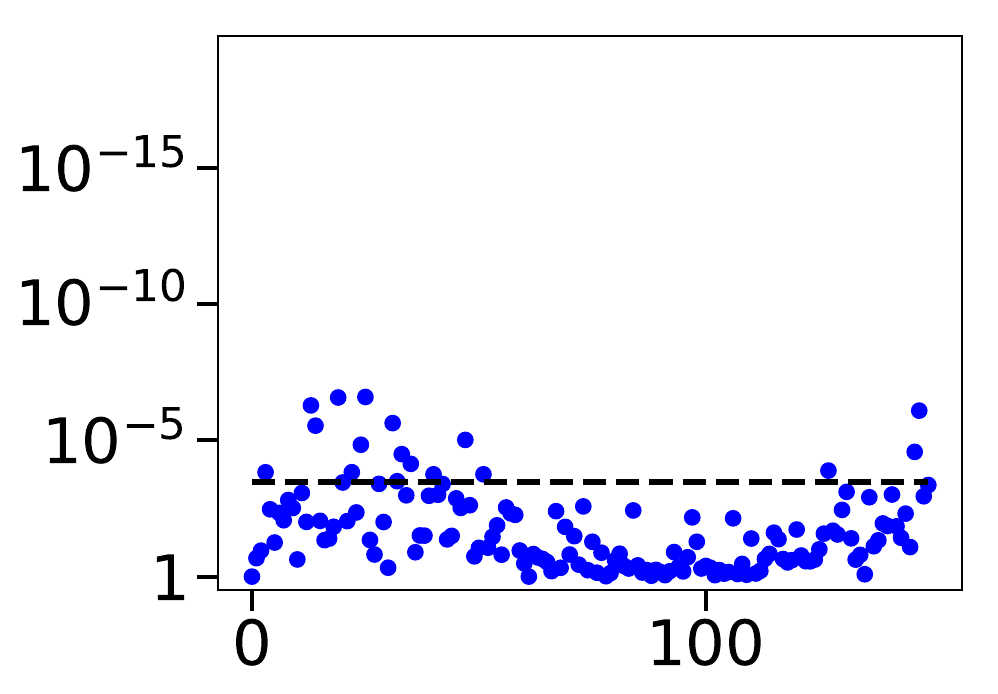}&
\includegraphics[width=\widthcq]{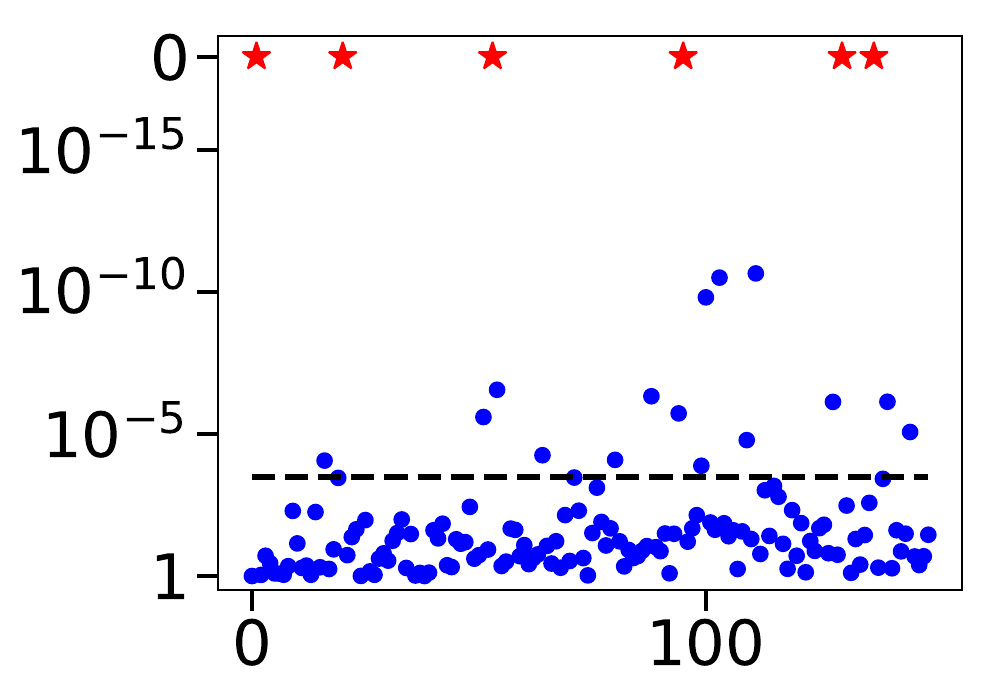}&
\includegraphics[width=\widthcq]{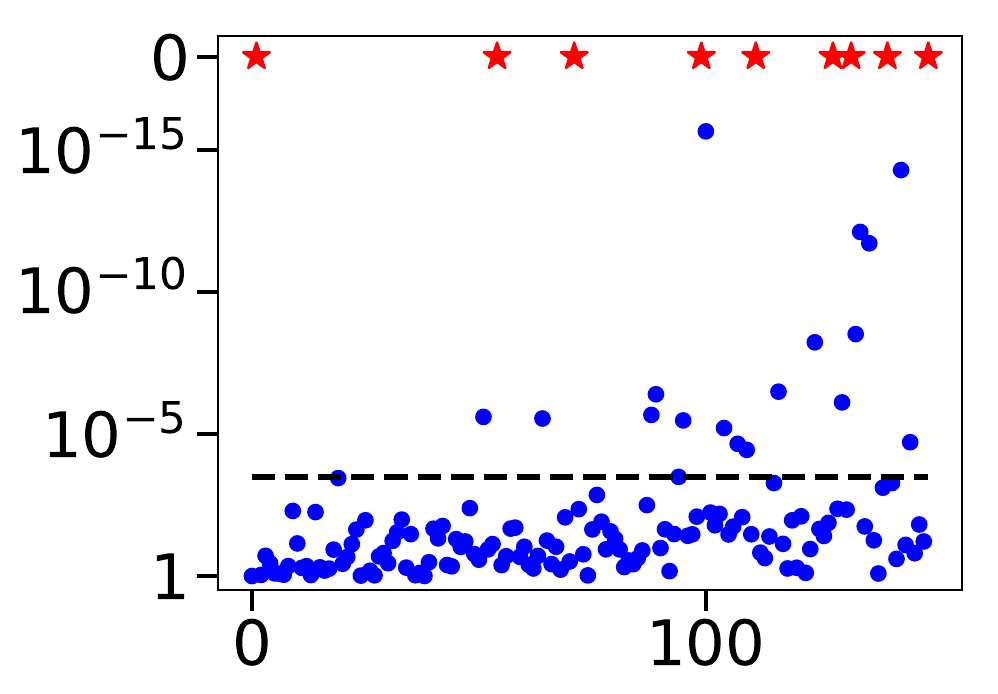}\\
\rotatebox{90}{\hspace{6em}\Rt}&
\includegraphics[width=\widthcq]{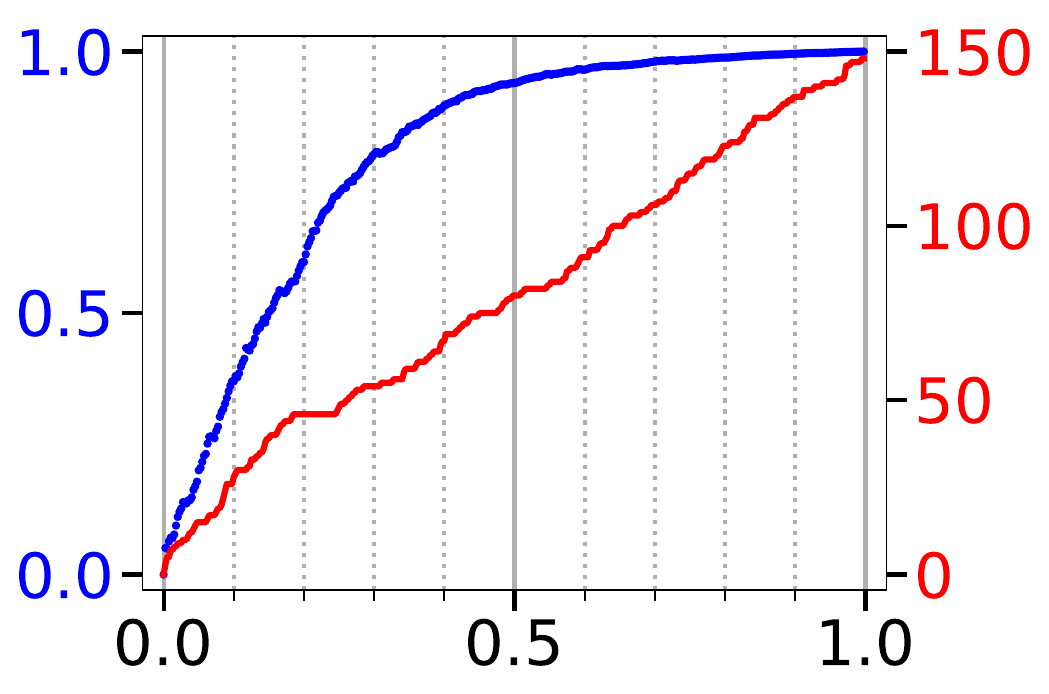}&
\includegraphics[width=\widthcq]{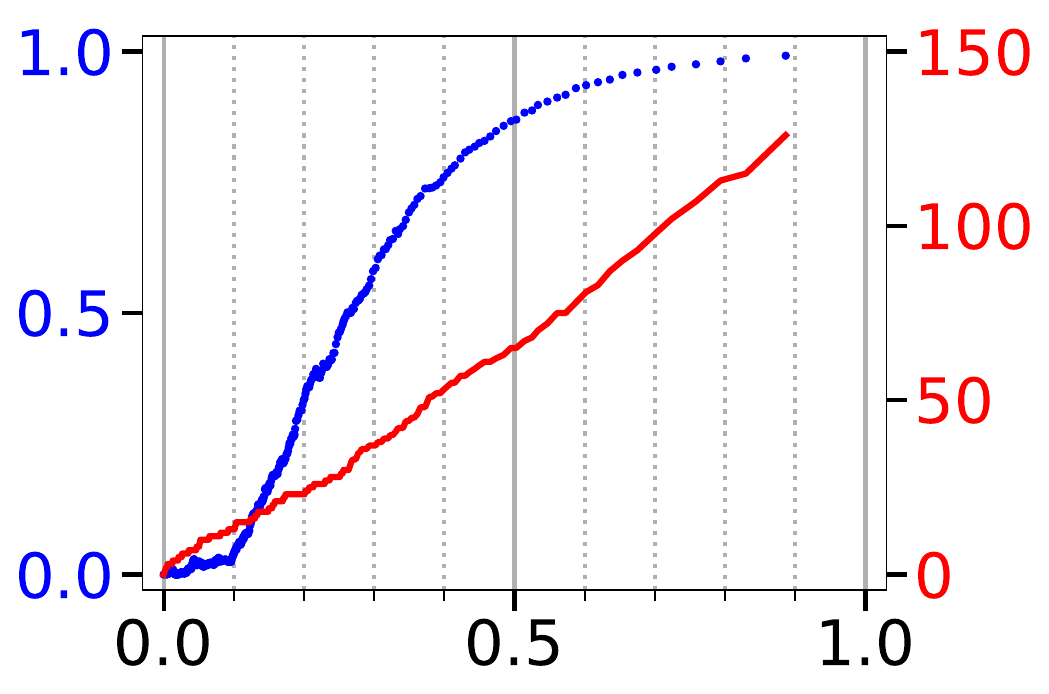}&
\includegraphics[width=\widthcq]{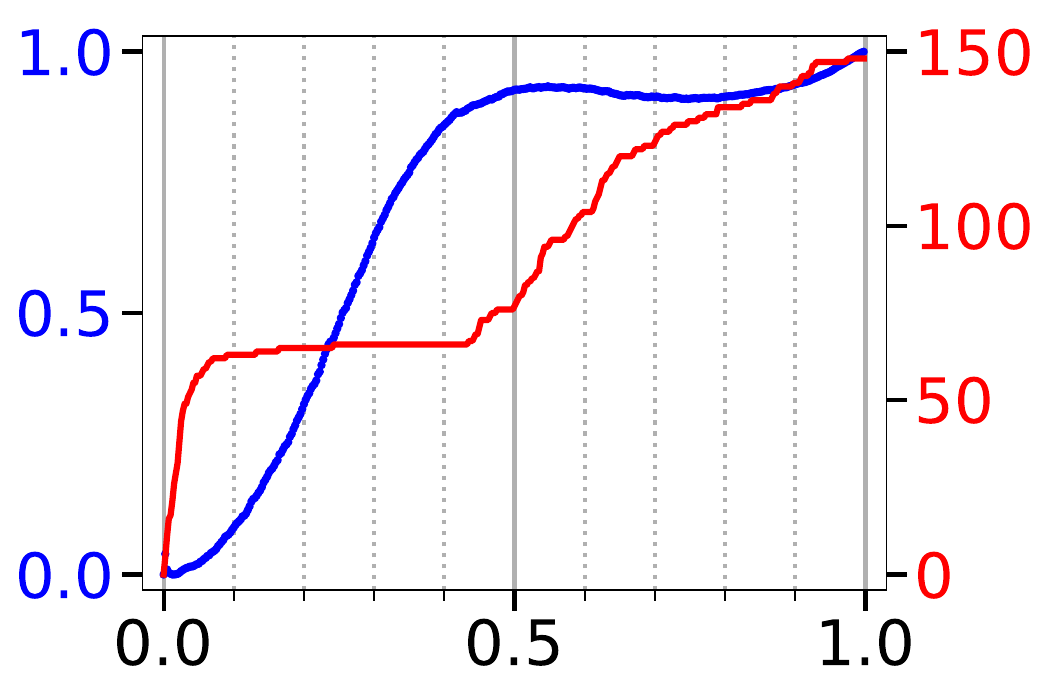}&
\includegraphics[width=\widthcq]{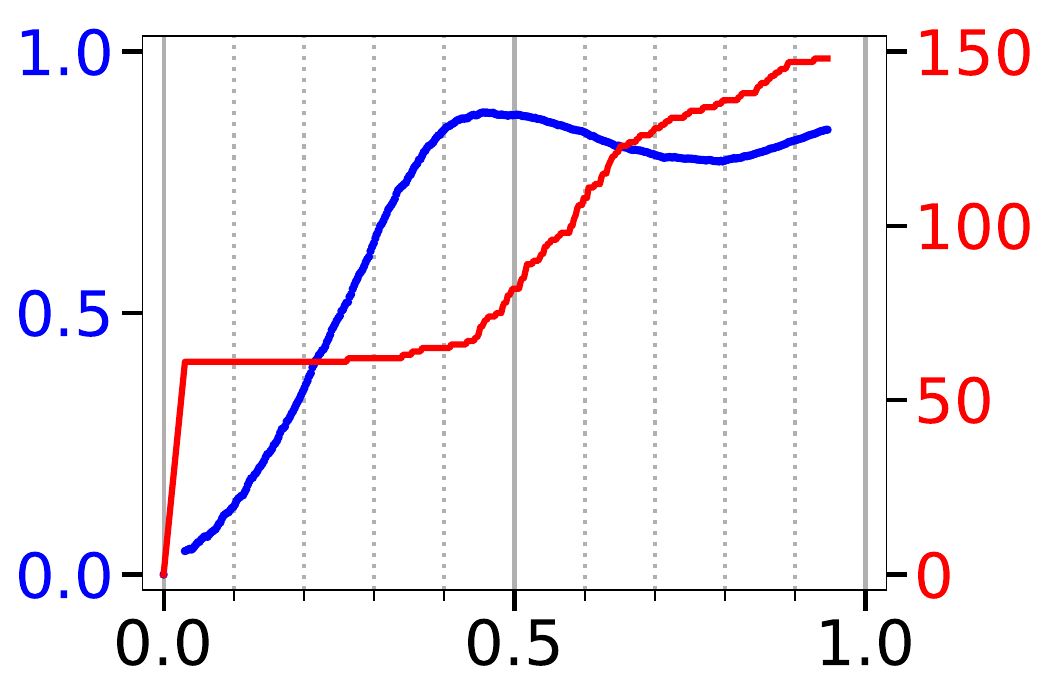}
\end{tabular}
\caption{\Lp\ provided accurate FDC than \ct\ and \si\ according to the histogram, KS, and \Rt\ tests on low-dimensional Geuvadis dataset. The layout is the same with \refig{reducednull}.\label{fig-reducedreal}}
\end{figure*}
Starting from the maximal DAG of the low-dimensional Geuvadis dataset, we first performed the overall histogram test on the results of \lp, \ct, and \si\ (\refig{reducedreal}, histogram). Compared to the low-dimensional null dataset (\refig{reducednull}), All three methods yielded an excess of small \pv{}s, reflecting the existence of genuine gene interactions. However, for \si, the excess was scarce at low $\lambda$, and overlapped with false positives at high $\lambda$. Consequently, \si\ had a low specificity on the Geuvadis data, which could not be resolved with finer choices of $\lambda$ (not shown). This also showed the difficulty in choosing the correct $\lambda$ for \si.

On the per-target evaluation with KS test, we excluded the genuine interactions by discarding the \pv{}s smaller than $0.01$. The remaining interactions were then compared against the uniform distribution $U(0.01,1)$. For \ct, we removed the bottom $1\%$ of all \pv{}s to account for its biased null distribution. Results were in agreement with those of the null dataset (\refig{reducedreal}, KS), confirmed the ideal FDC from \lp, and remained stable at higher exclusion thresholds (not shown).

\begin{table}
\center
\caption{\Lp\ outperformed \ct\ and \si\ on AUR2 (\refig{reducedreal}) of the low-dimensional Geuvadis dataset. The layout is the same with \reftab{1}.\label{tab-reducedreal-auc}}\vspace{1em}
\begin{tabular}{c|c|c|cc}
&\lp&\ct&\multicolumn{2}{c}{\si}\\
Bound&&&$\lambda=0.001$&0.02\\
\hline
$[0,0.01]$&\textbf{0.049}&0.003&0.013&0.007\\
$[0,0.05]$&\textbf{0.11}&0.008&0.010&0.034\\
$[0,0.2]$&\textbf{0.35}&0.094&0.11&0.15\\
$[0,1]$&\textbf{0.81}&0.70&0.70&0.66
\end{tabular}
\end{table}
The existence of genuine interactions violated null linearity and lowered $R^2$ for all three methods, more so on the leading method \lp\ (\refig{reducedreal}, \Rt). This was also shown in the AUR2 of these methods (\reftab{reducedreal-auc}). As expected, the violation appeared small and localized at small \pv{}s and low recall, because the number of genuine regulators is low for every target gene. Other than that, method performances mostly agreed with those on the low-dimensional null dataset, including that \si\ continued to assign highly significant regulators to a small number of targets.

In summary, the performances of all three methods on the low-dimensional Geuvadis dataset agreed highly with those on the low-dimensional null dataset. This has several implications. First, the statistical tests of FDC in \refssec{fdr} could also be applied on real gene regulation networks, after adjustments for sparse interactions. Second, the null dataset was validated to highly resemble the null interactions in the Geuvadis dataset. This supported our upcoming FDC evaluations with the simulated high-dimensional null dataset. Third, we continued to find \lp\ as the best method for FDC.

\subsection{Lassopv obtained accurate false discovery control on the high-dimensional datasets}
In high-dimensional problems, lasso regression becomes under-determined as the regularization strength approaches zero. Machine precision prevents some of the insignificant regulators from having any role in the regression, and therefore biases their \pv{}s to one. This challenge can only be overcome by a high-precision lasso solver. However, this would not affect our analysis of identifying significant regulations, because we are only interested in the small \pv{}s.

\begin{figure}[!tpb]
\center
\begin{tabular}{p{0em}p{1.03\widthcq}p{0em}p{0.93\widthcq}}
\vspace{0em}\textbf{A}&\vspace{0em}\includegraphics[width=\linewidth]{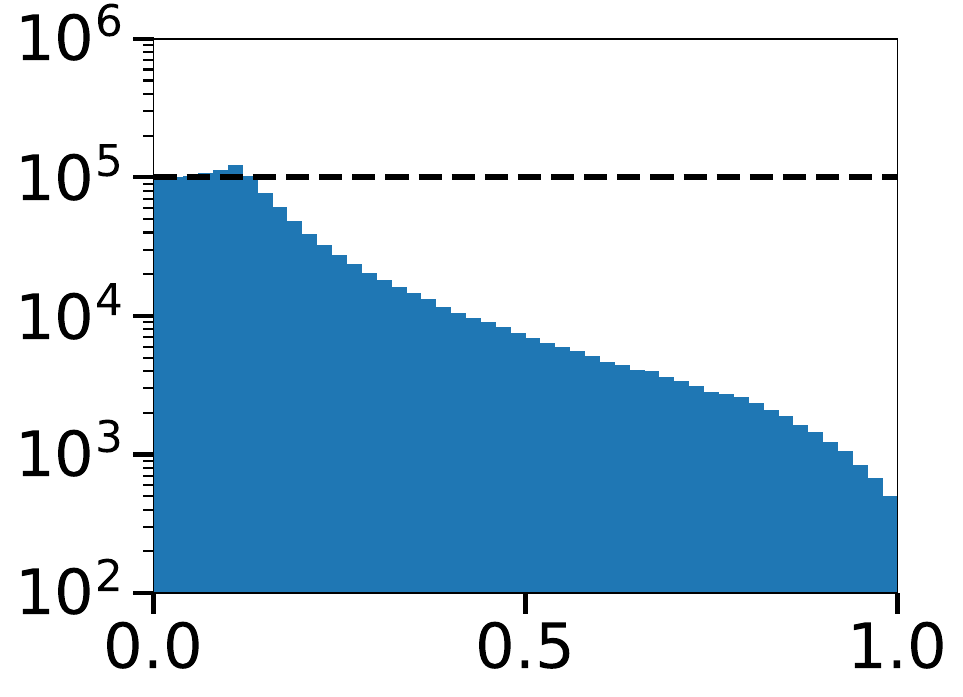}&
\vspace{0em}\textbf{D}&\vspace{0em}\includegraphics[width=\linewidth]{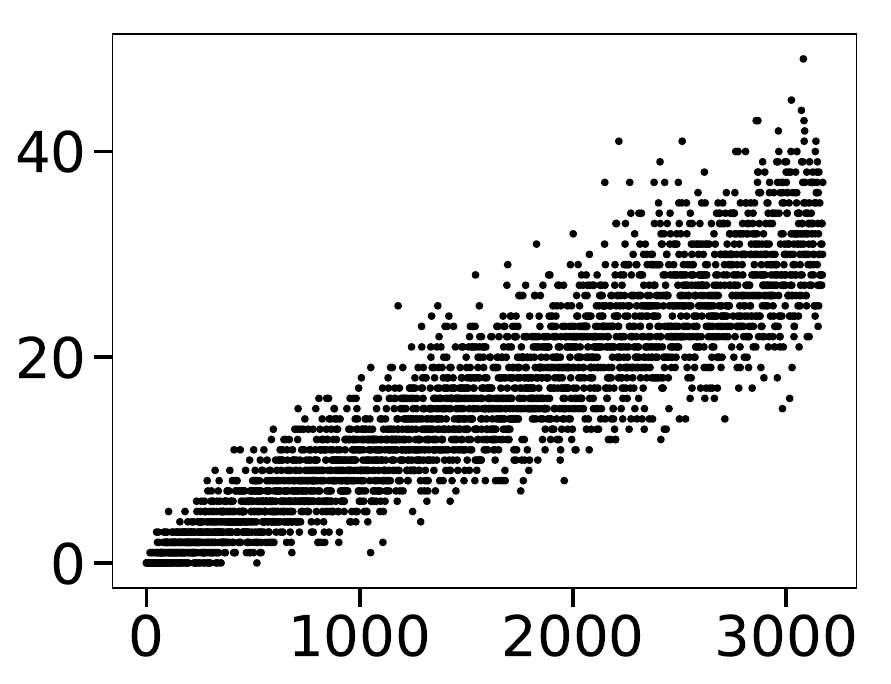}\\
\vspace{0em}\textbf{B}&\vspace{0em}\includegraphics[width=\linewidth]{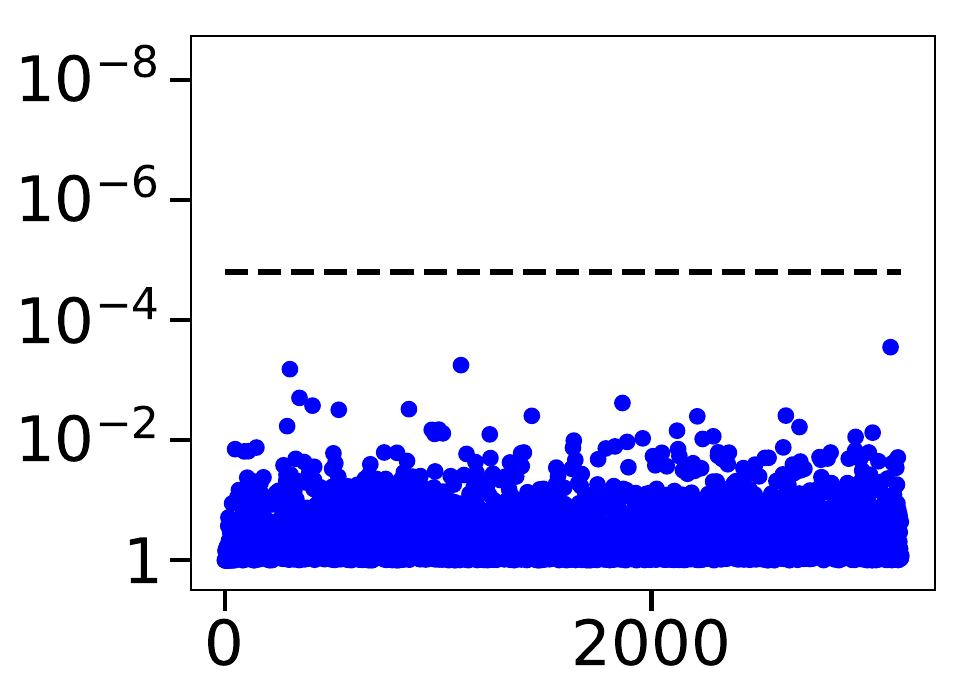}&
\vspace{0em}\textbf{E}&\vspace{0em}\includegraphics[width=\linewidth]{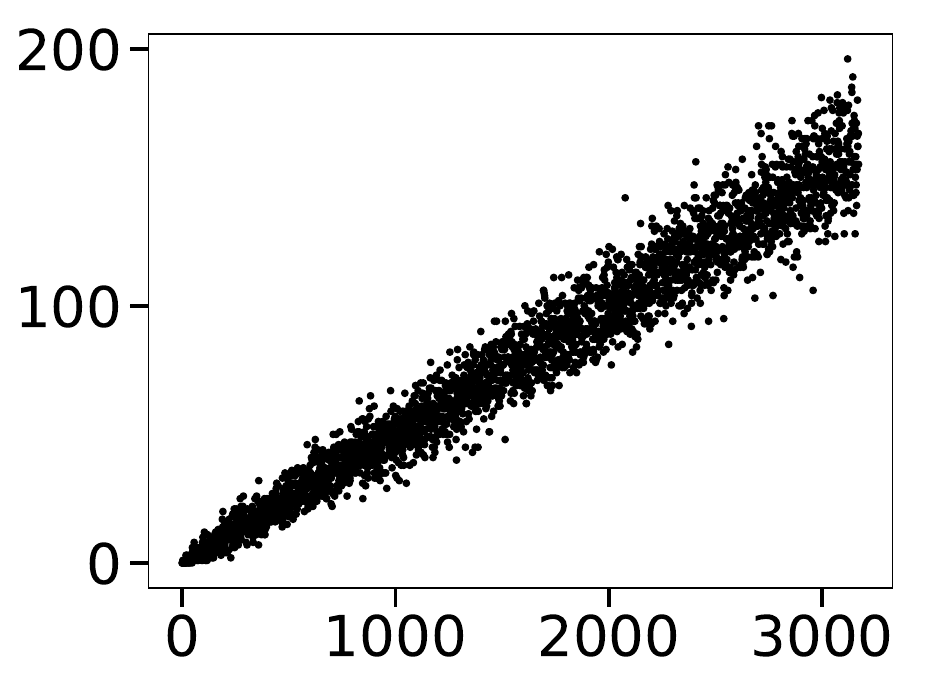}\\
\vspace{0em}\textbf{C}&\vspace{0em}\includegraphics[width=\linewidth]{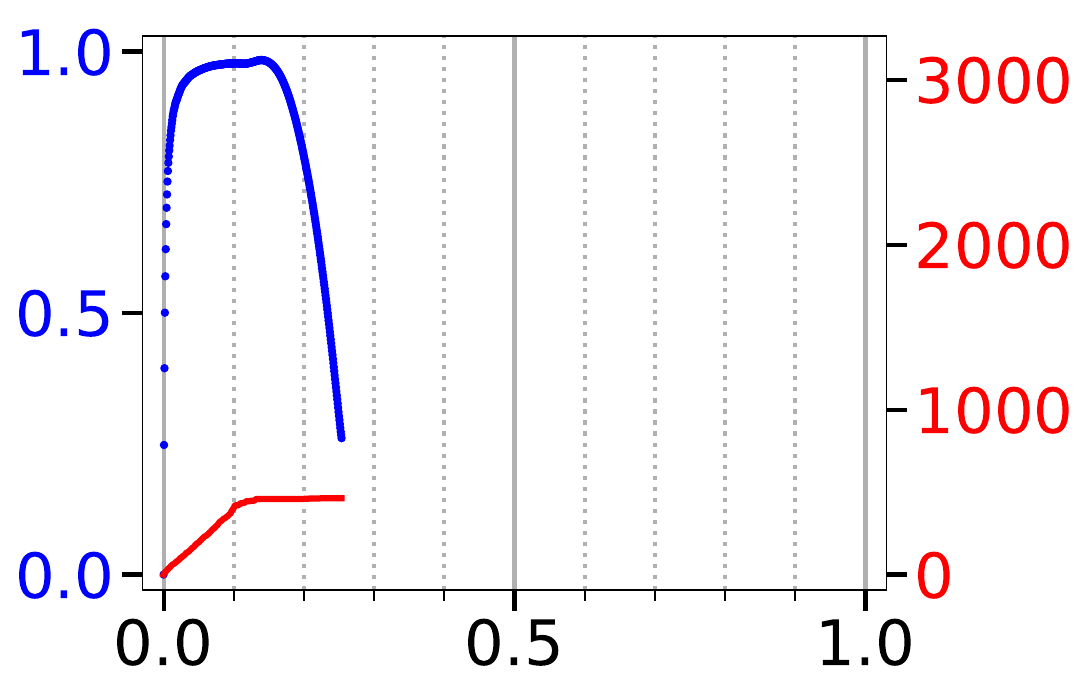}&
\vspace{0em}\textbf{F}&\vspace{0em}\includegraphics[width=\linewidth]{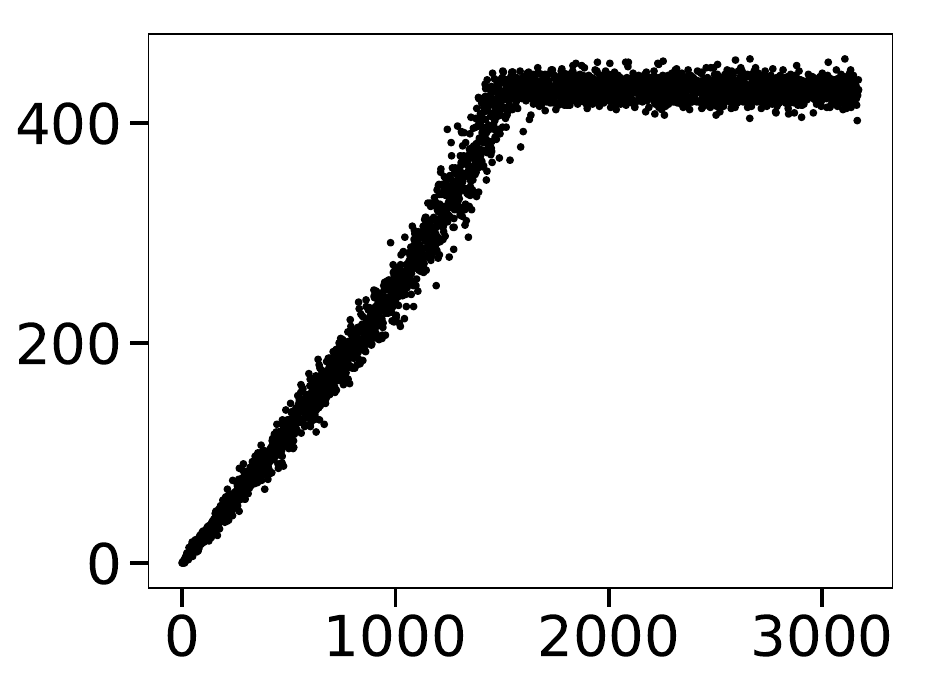}
\end{tabular}
\caption{\Lp\ provided accurate FDC on the high-dimensional null dataset. (\textbf{A, B, C}) The histogram (\textbf{A}), KS (\textbf{B}), and \Rt\ (\textbf{C}) tests following the same layout as \refig{reducednull}. To account for high dimensionality, we removed \pv{}s=1 in (\textbf{A}), and only tested the bottom 5\% \pv{}s in (\textbf{B}). \textbf{(D, E, F)} The linearity test for \lp{} between the numbers of predictors ($x$) and significant predictors ($y$) at \pv\ significance thresholds of the bottom 1\% (\textbf{D}, $R^2=0.84$), 5\% (\textbf{E}, $R^2=0.96$), and 20\% (\textbf{F}, $R^2=0.80$) potential regulations. Every dot corresponds to one variable selection in the network inference.\label{fig-fullnulllasso}}
\end{figure}

For that reason, we tested \lp\ and \ct\ directly on the high-dimensional null dataset containing 3172 genes. \Si\ could not handle high-dimensional scenarios and therefore was excluded from the analysis. As confirmed for \lp\ in \refig{fullnulllasso}\textbf{A}, the high-dimensional effect unavoidably biased a large proportion of insignificant regulations to \pv=1 (removed in figure). The resulting \pv{} distribution differed notably from the standard uniform distribution, but only for insignificant regulations (\pv$>$0.1). In spite of that, its FDC remained accurate because \pv{}s$<$0.1 were still uniformly distributed. The slight over-abundance of \pv{}s between 0.1 and 0.2 was due partly to the same high-dimensional effect, and partly to the analytical approximation in \refsec{pv}. A simulation-based \pv\ computation without the approximation shrank the peak by half (not shown).

\begin{table}
\center
\caption{\Lp\ outperformed \ct\ and \si\ on AUR2 (\refig{fullnulllasso}\textbf{C}, \refig{fullnullcovtest}, \refig{fullreal}) of the high-dimensional null and Geuvadis datasets. Best performers are shown in bold.\label{tab-aur2-full}}\vspace{1em}
\begin{tabular}{c|cc|cc}
&\multicolumn{2}{c|}{\textbf{Null dataset}}&\multicolumn{2}{c}{\textbf{Geuvadis dataset}}\\
Bound&\lp&\ct&\lp&\ct\\
\hline
$[0,0.01]$&\textbf{0.65}&0.008&0.01&\textbf{0.06}\\
$[0,0.05]$&\textbf{0.88}&0.10&\textbf{0.31}&0.06\\
$[0,0.2]$&\textbf{0.94}&0.08&\textbf{0.75}&0.06\\
\end{tabular}
\end{table}

\begin{figure}
\center
\begin{tabular}{p{0em}p{1.03\widthcq}p{0em}p{0.93\widthcq}}
\vspace{0em}\textbf{A}&\vspace{0em}\includegraphics[width=\linewidth]{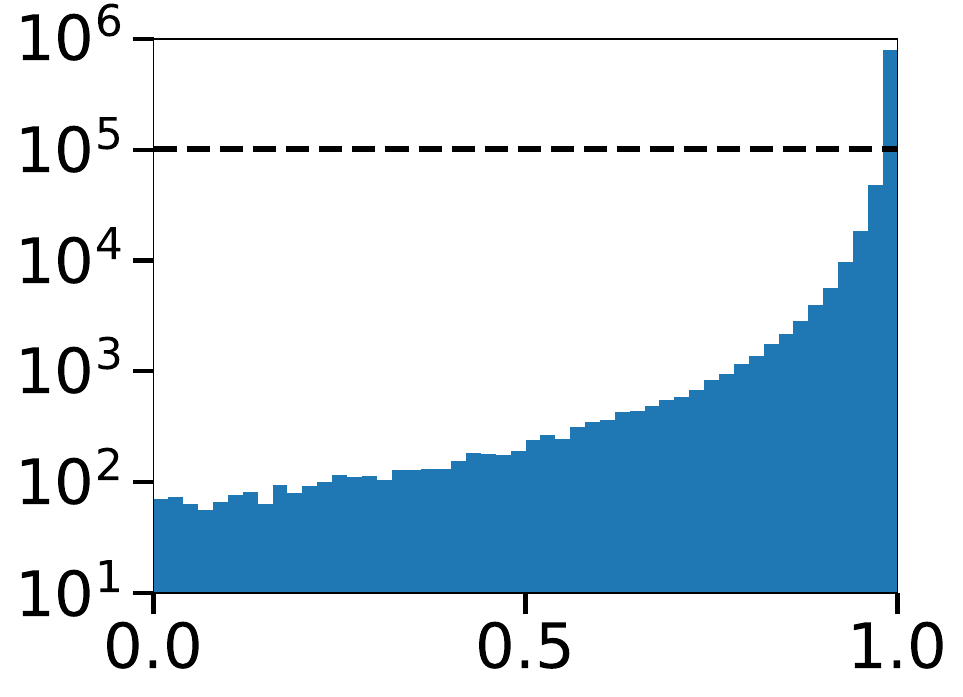}&
\vspace{0em}\textbf{D}&\vspace{0em}\includegraphics[width=\linewidth]{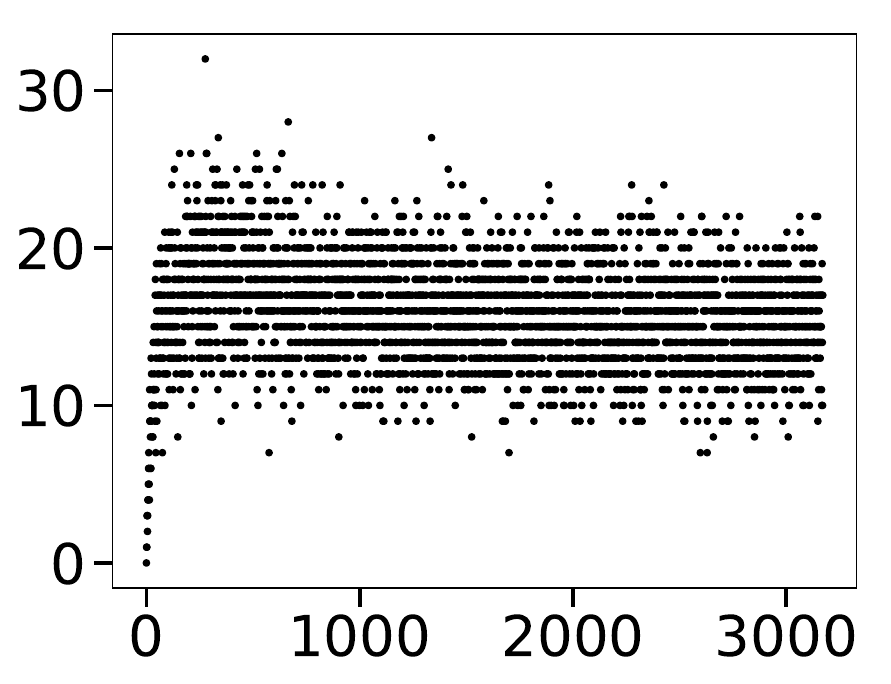}\\
\vspace{0em}\textbf{B}&\vspace{0em}\includegraphics[width=\linewidth]{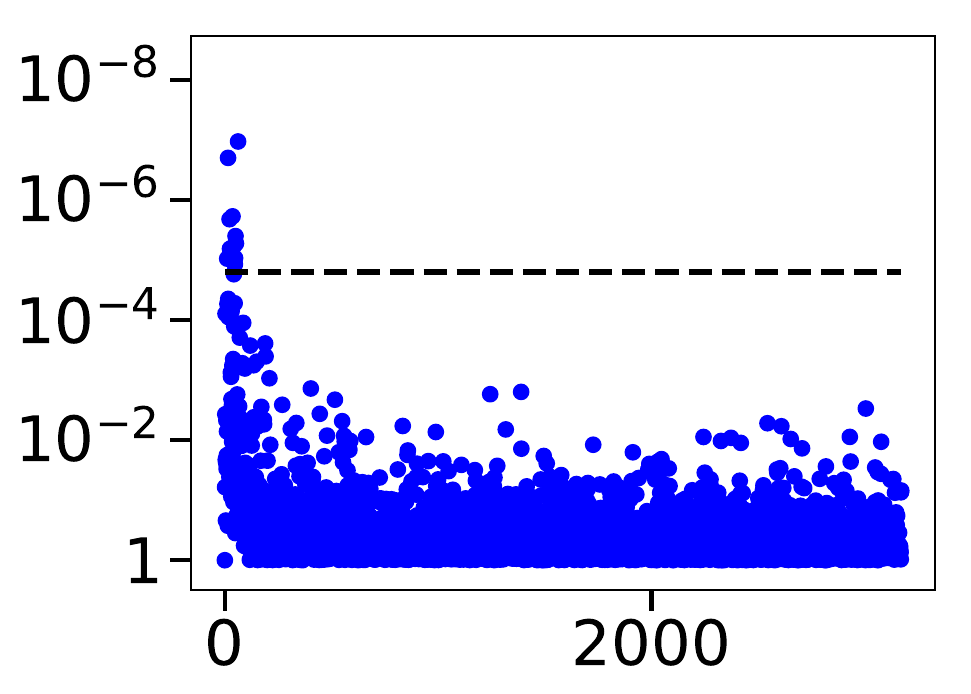}&
\vspace{0em}\textbf{E}&\vspace{0em}\includegraphics[width=\linewidth]{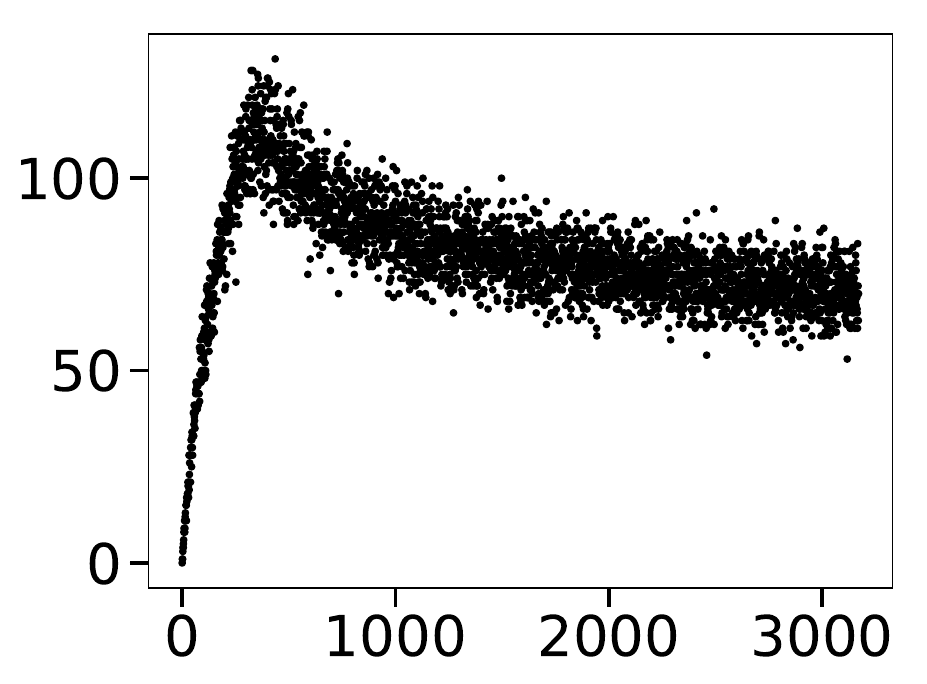}\\
\vspace{0em}\textbf{C}&\vspace{0em}\includegraphics[width=\linewidth]{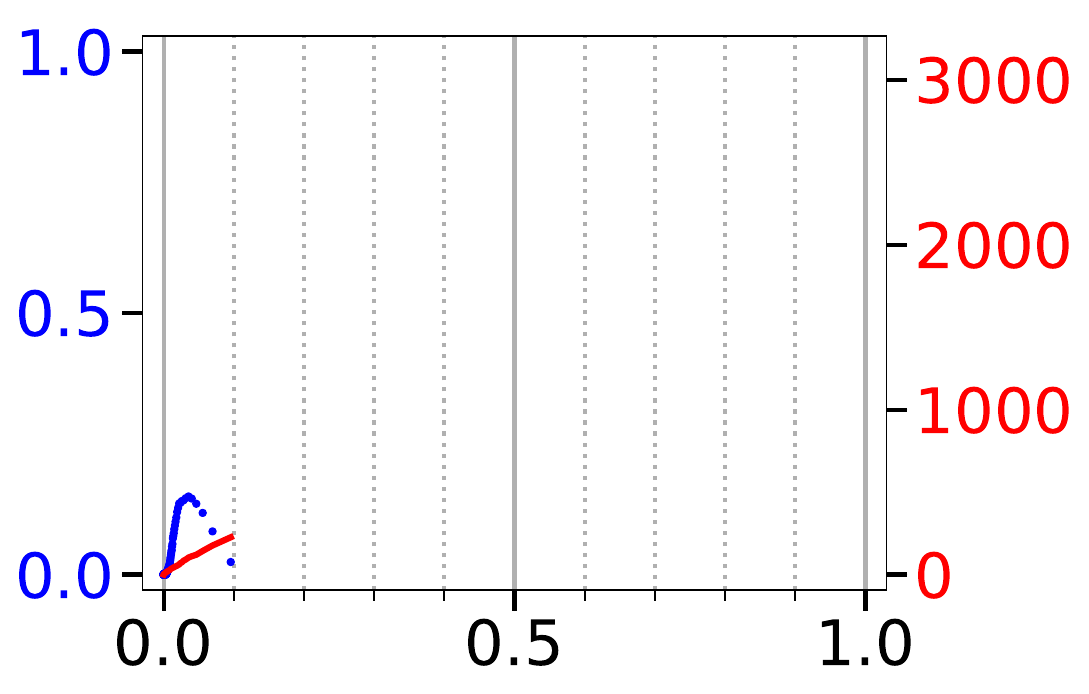}&
\vspace{0em}\textbf{F}&\vspace{0em}\includegraphics[width=\linewidth]{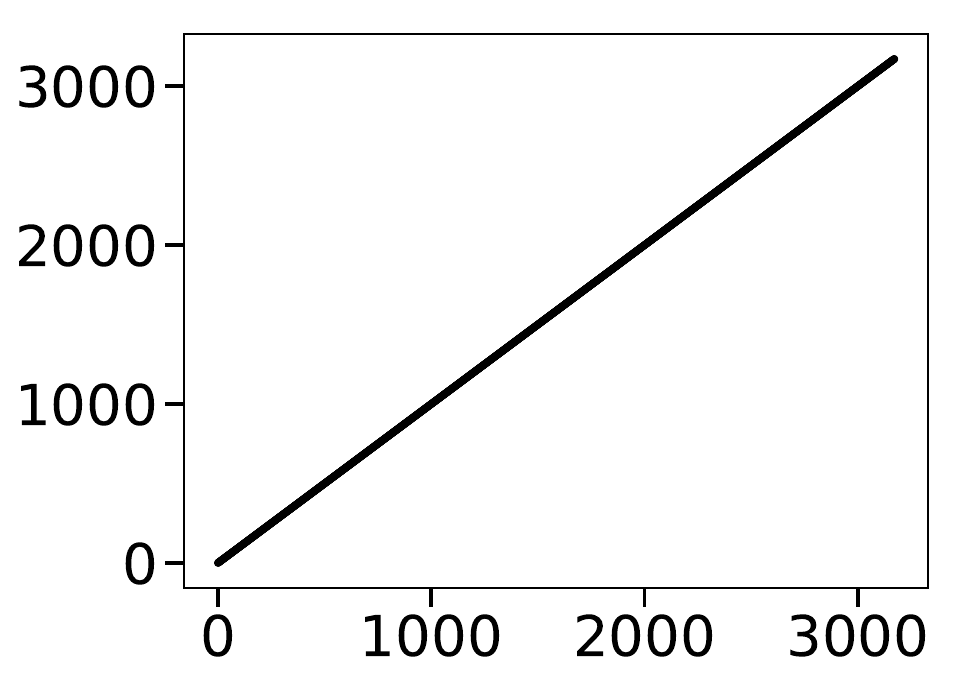}
\end{tabular}
\caption{\ct\ failed to provide FDC on the high-dimensional null dataset. The layout is the same with \refig{fullnulllasso}. (\textbf{B}) the KS test was performed against the empirical distribution of all \pv{}s (of the bottom 5\%). (\textbf{D, E}) \ct{} could not provide FDC in network inference by failing the linearity test, at \pv\ significance thresholds of the bottom 1\% (\textbf{D}, $R^2=0.03$) and 5\% (\textbf{E}, $R^2=0.13$). (\textbf{F}) \ct\ claimed all gene regulations significant at the bottom 20\% threshold due to ties.\label{fig-fullnullcovtest}}
\end{figure}

\begin{figure}
\center
\begin{tabular}{lcc}
&\lp&\ct\\
\rotatebox{90}{\hspace{4em}Histogram}&
\includegraphics[width=\widthcq]{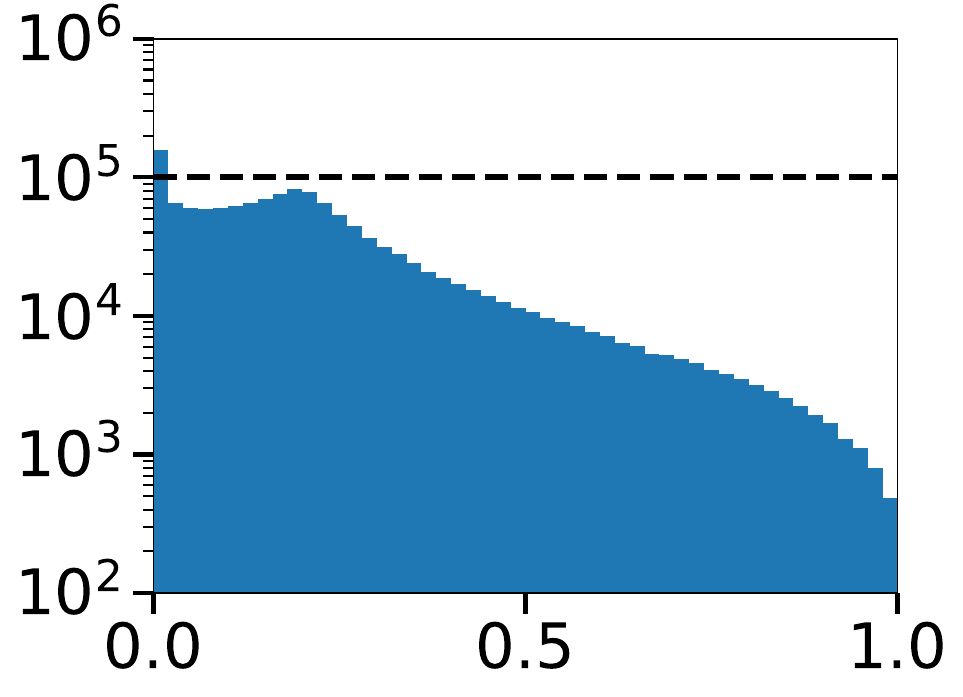}&
\includegraphics[width=\widthcq]{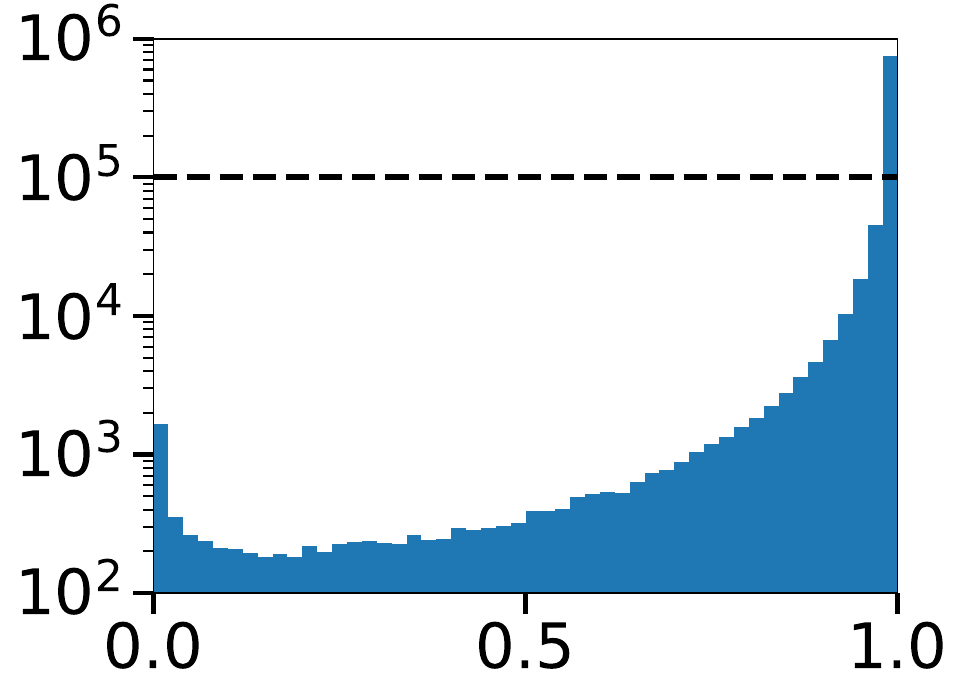}\\
\rotatebox{90}{\hspace{6em}KS}&
\includegraphics[width=\widthcq]{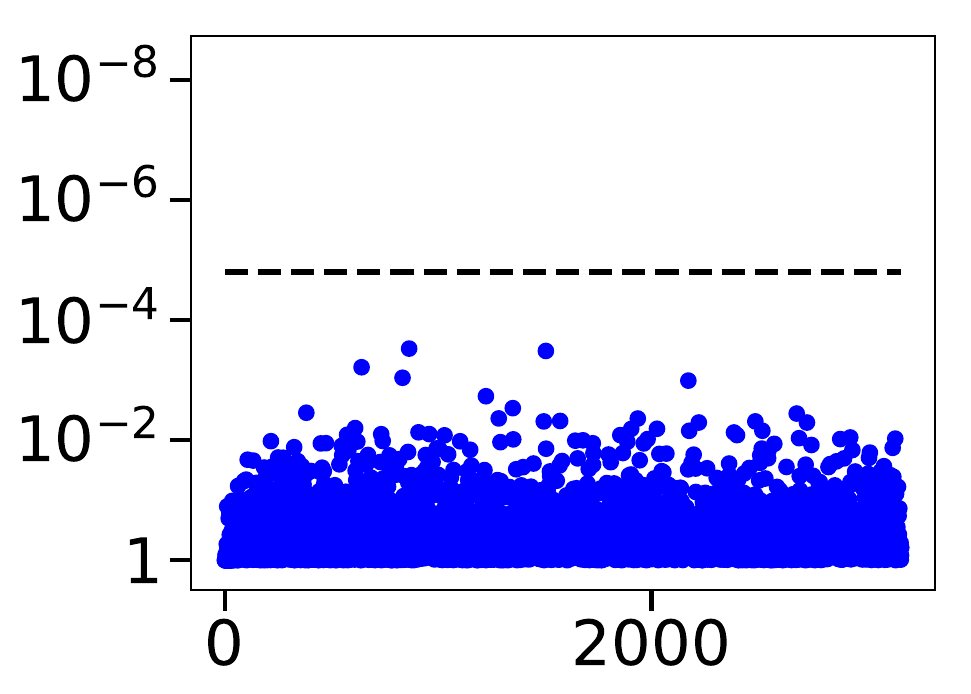}&
\includegraphics[width=\widthcq]{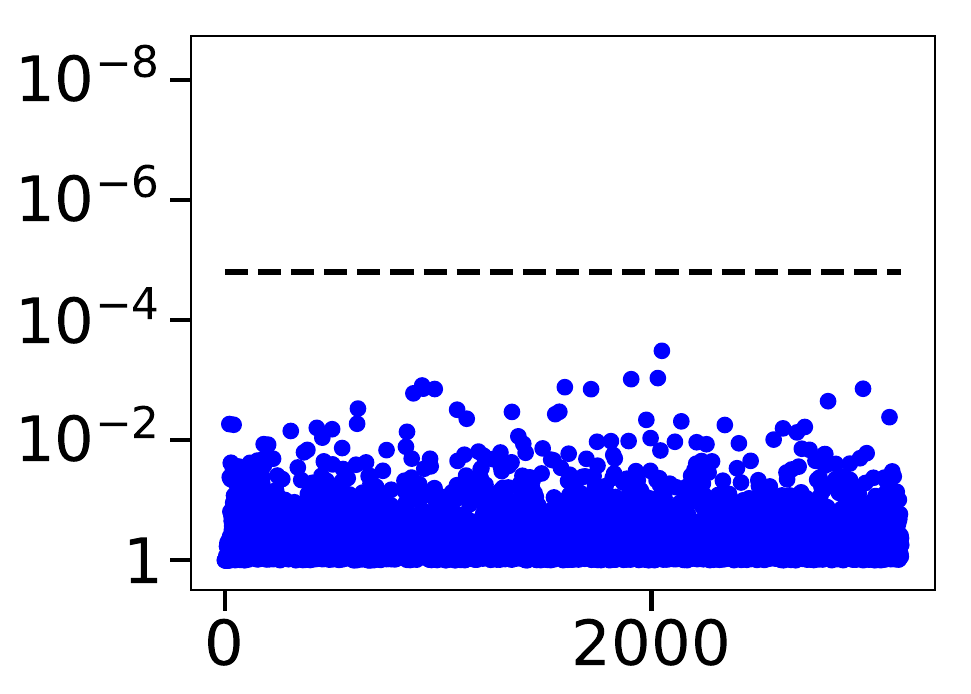}\\
\rotatebox{90}{\hspace{6em}\Rt}&
\includegraphics[width=\widthcq]{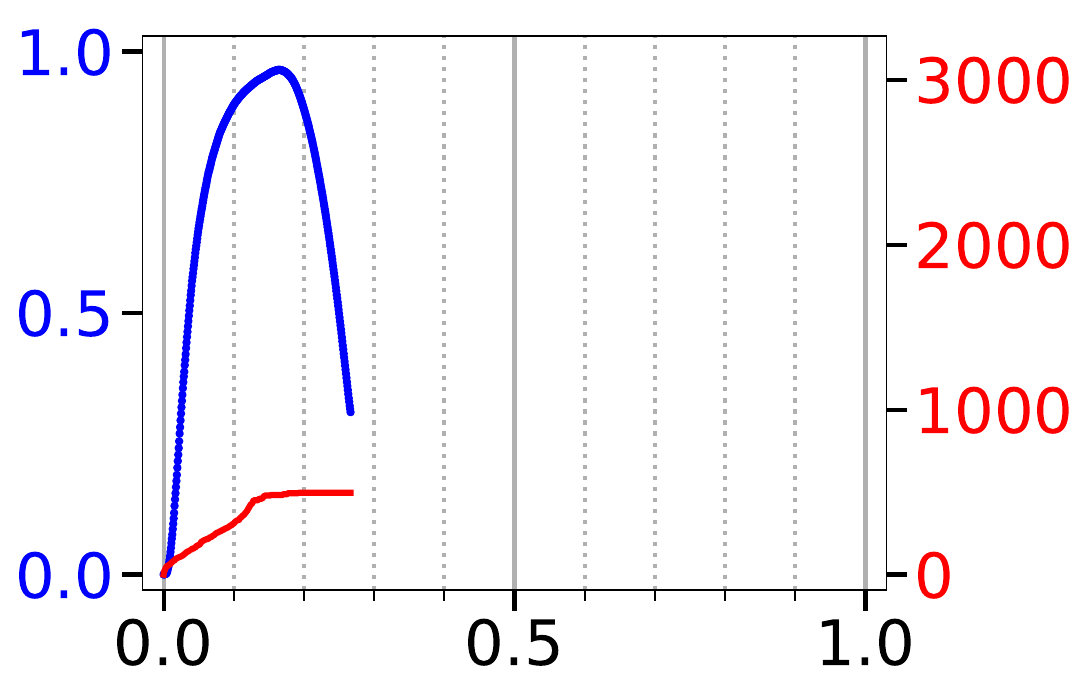}&
\includegraphics[width=\widthcq]{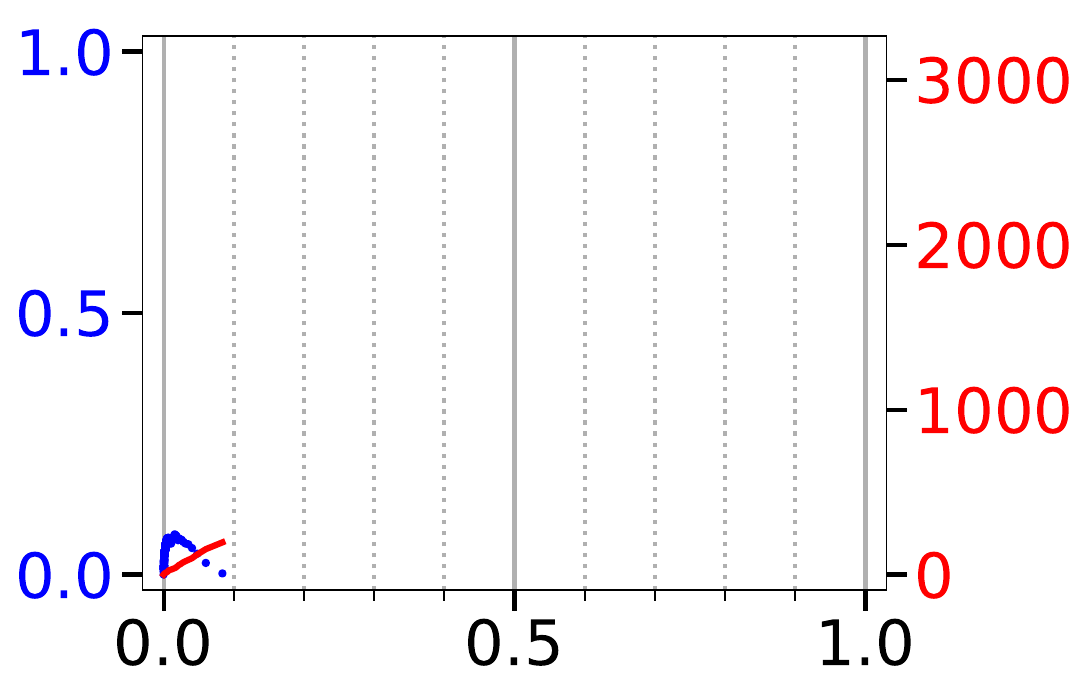}
\end{tabular}
\caption{\Lp\ provided accurate FDC than \ct\ on the high-dimensional Geuvadis dataset. The layout is the same with \refig{reducednull}. Besides, to remove the effects of high dimensionality and geuine interactions, only the bottom 3\% to 5\% (totalling 2\%) of all \pv{}s were evaluated in the KS tests.\label{fig-fullreal}}
\end{figure}
\Lp\ also displayed accurate FDC in other tests. The KS test on the bottom 5\% \pv{}s confirmed \lp's accurate FDC on a per variable selection basis (\refig{fullnulllasso}\textbf{B}). In network inference, \lp\ attained a highly linear relation ($R^2=0.96$) between the numbers of potential and significant regulators (\refig{fullnulllasso}\textbf{DE}), as a typical example of what a FDC conforming network reconstruction method should yield in the linearity test. Although the high-dimensional effect resulted in plateaus at weak thresholds (\refig{fullnulllasso}\textbf{CF}), these again did not affect FDC at small \pv{}s. Its proper network FDC was also demonstrated in the \Rt\ curve (\refig{fullnulllasso}\textbf{C}) and AUR2 (\reftab{aur2-full}).

On the other hand, \ct\ could not achieve FDC in high-dimensional network inference. Particularly, the nonlinear relation between the numbers of potential and significant regulators indicated a lack of FDC (\refig{fullnullcovtest}\textbf{DE}). This was also manifested in its failed KS and \Rt\ tests (\refig{fullnullcovtest}\textbf{BC}, \reftab{aur2-full}). Besides that, the high-dimensional effect biased \ct's insignificant regulations towards \pv=1 even more strongly than \lp's (\refig{fullnullcovtest}\textbf{AF}). The same conclusions hold for \lp\ and \ct\ with the high-dimensional Geuvadis dataset. After accounting for the genuine interactions, \lp\ again showed optimal FDC (\refig{fullreal}, \reftab{aur2-full}).

Although the high dimensionality of modern datasets introduces unavoidable distortions in insignificant \pv{}s, and the groundtruth is unavailable, we found that \lp\ could still provide ideal FDC for small \pv{}s, and therefore is unaffected by high-dimensional effects. We established the principles and tests to evalute FDC in network inference through \lp's symbolic linearity between the numbers of candidate and significant regulators. \Lp\ remained the optimal FDC method for high-dimensional network inference.

\subsection{Lassopv reduced false discovery by removing the spurious indirect regulations in Pearson co-expression network}
Hardly any software package on the market is capable of FDC in a high-dimensional network reconstruction problem. To our knowledge, the only available ones are co-expression networks, except the \ct\ which is already shown highly biased. However, co-expression networks are well known for their high FPRs from indirectly regulated and confounded genes, due to its pairwise nature. Regression based methods can reduce these false positives by accounting for the direct and indirect effects together.

\begin{figure}[!tpb]
\center
\begin{tabular}{p{0em}p{0.94\widthcq}p{0em}p{1.01\widthcq}}
\vspace{0em}\textbf{A}&\vspace{0em}\includegraphics[width=\linewidth]{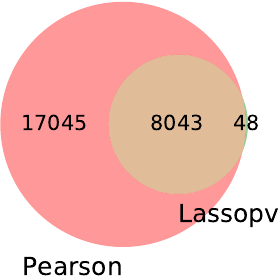}&
\vspace{0em}\textbf{F}&\vspace{0em}\includegraphics[width=0.2\linewidth]{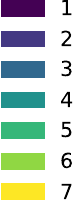}
\\
\vspace{0em}\textbf{B}&\vspace{0em}\includegraphics[width=\linewidth]{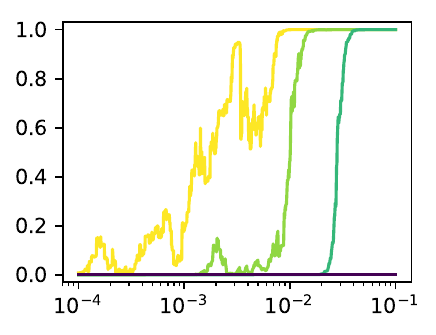}&
\vspace{0em}\textbf{C}&\vspace{0em}\includegraphics[width=\linewidth]{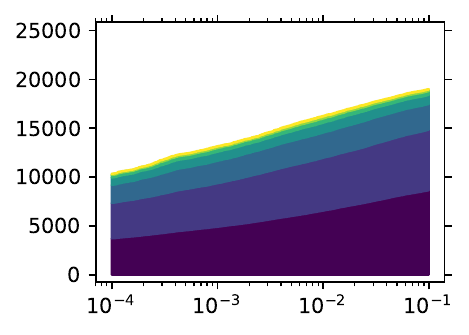}\\
\vspace{0em}\textbf{D}&\vspace{0em}\includegraphics[width=\linewidth]{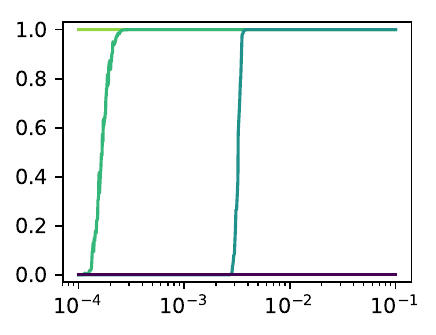}&
\vspace{0em}\textbf{E}&\vspace{0em}\includegraphics[width=\linewidth]{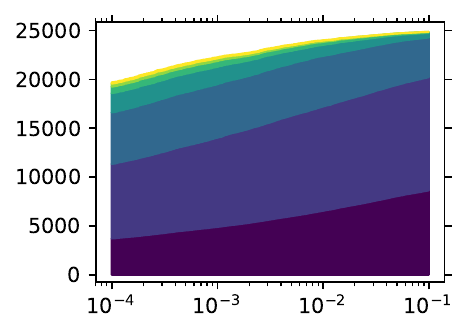}
\end{tabular}
\caption{\Lp\ reduced the false positives in co-expression networks. (\textbf{A}) Venn diagram of regulation overlap. (\textbf{B-E}) Hypergeometric enrichment \pv{}s (\textbf{BD}) and the total numbers (\textbf{CE}) of significant co-expressions ($y$) in the \lp\ networks of indirect regulations (\textbf{BC}) and indirect regulations + confounding relations (\textbf{DE}) of a given step, at different significance thresholds of Bonferroni adjusted \pv\ ($x$). (\textbf{F}) Color legend for \textbf{B-E}. Colors represent different numbers of direct steps in indirect or confounding relations, where one step is the inferred network of direct regulations.\label{fig-pearson}}
\end{figure}
We tested the reduction of false positives from \lp, by comparing it with Pearson co-expression network on the high-dimensional Geuvadis dataset with 3172 genes. Using the same significance threshold of $0.05$ with Bonferroni corrections across all interactions, \lp\ discovered 8091 regulations as opposed to 25088 in the co-expression network, which mostly overlap (\refig{pearson}\textbf{A}). We derived indirect regulations of the \lp\ network by connecting steps of significant direct regulations. Indirect regulations up to 4-step far were enriched with significant co-expressions (\refig{pearson}\textbf{B}), and covered over 70\% of all co-expressions at the adjusted \pv{} cutoff $0.05$ (\refig{pearson}\textbf{C}). Further inclusion of confounded genes led to a much denser network, with an enrichment of co-expressions only up to 3-step far (\refig{pearson}\textbf{D}), but covering around 95\% of all co-expressions (\refig{pearson}\textbf{E}).

Although a high-quality gold standard is not available on real data, we still found convincing evidence supporting the reduction of indirect and confounding false positives by Bayesian networks, particularly \lp, from co-expression networks.

\section{Conclusion}
In this article, we presented the problem of false discovery control in Bayesian network reconstruction and its impact on inference accuracy and network sparsity. We designed four statistical tests to evaluate the FDC of reconstructed networks on real genomic data. We proposed a new method to repurpose lasso regression for variable selection and computed its \pv{} in the program \lp\ for consistent FDC in network inference. On simulated and real, low- and high-dimensional genomic datasets, our statistical tests revealed hidden defects in \ct\ and \si --- the existing variable selection methods also from lasso regression --- and demonstrated the advantageous, accurate \pv{}s and FDC from \lp.

Already known in existing GWAS's, consistent FDC is crucial for optimal prediction accuracy. Consequently, FDC evaluations can advise and guide the application and design of network inference methods, and should be performed on all reconstructed Bayesian networks of any continuous value or merely a list of interactions, although this paper only evaluated different \pv{}s. On the other hand, as a hypothesis testing method for variable selection, \lp\ is potentially applicable, for example, on causal (expression) quantitative trait loci discovery and supervised dimensional reduction, and also on other disciplines beyond computational biology.

This paper is not concerned with how to obtain a prior ordering of gene nodes. We used genotype and gene expression data of the same individuals in population-based studies to orient the causal direction between any pair of genes \cite{rockman2008reverse,li2010critical} and obtained a dense prior DAG. Depending on data availability, the prior ordering may also be obtained from known regulatory gene function annotations or regulatory interactions (e.g. protein-DNA interactions) \cite{Shojaie:2010}, or simply from a greedy maximum-likelihood optimization in the absence of any additional information \cite{schmidt:2007}.

Similarly, this paper only reconstructs the structure of the Bayesian network. Since the Bayesian network reflects the conditional dependency between gene expression levels, a predictive model may be derived with an additional regression of any target gene on its parental genes in the reconstructed network. Under a linear approximation for the sparse regulatory system, an unpenalized linear regression may suffice.

\section*{Funding}
The authors gratefully acknowledge the support by BBSRC (grant numbers BB/P013732/1 and BB/M020053/1).\vspace*{-12pt}

\bibliographystyle{unsrt}

\bibliography{bib,bib2}

\clearpage
\renewcommand\thefigure{S\arabic{figure}}
\renewcommand\thetable{S\arabic{table}}
\renewcommand\thesection{S\arabic{section}}
\renewcommand\theequation{S\arabic{equation}}
\renewcommand\thepage{S\arabic{page}}
\setcounter{figure}{0}
\setcounter{table}{0}
\setcounter{section}{0}
\setcounter{equation}{0}
\setcounter{page}{1}

\begin{center}
\Large Supplementary Information
\end{center}
\section{Computation of lasso \pv{}s}
\label{sec-pv}
To compute the null distribution for variable $\vec x_i$, we simply assume the null hypothesis $H_0^{(i)}$ holds and omit its conditioning. The \pv\ expression in \refeq{pv-s} can be decomposed into:
\eqa{&&P\left(\lambda_i\ge\tilde\lambda_i\right)\nonumber\\
&=&P\left(\sup\{\lambda:\hat\beta_i(\lambda)\ne0\}\ge\tilde\lambda_i\right)\nonumber\\
&=&P\left(\hat\beta_i(\tilde\lambda_i)\ne0)\right)\nonumber\\
&&+P\left(\sup\{\lambda:\hat\beta_i(\lambda)\ne0\}\ge\tilde\lambda_i\wedge\hat\beta_i(\tilde\lambda_i)=0)\right).\label{eq-lasso-psplit}}
The first term is the probability that $\vec x_i$ is \textit{active} (i.e.\ $\hat\beta_i\ne 0$ by definition) at $\lambda=\tilde\lambda_i$, whilst the second is the probability that $\vec x_i$ is active at some $\lambda>\tilde\lambda_i$ but inactive at $\lambda=\tilde\lambda_i$.

For every predictor, starting and stoping to be active are also call \textit{entering and leaving the active set}, whose critical $\lambda$ values are named \textit{knots}. In lasso regression, predictors can enter and leave the active set multiple times. Therefore the second term in \refeq{lasso-psplit} is nonzero in general. Here we make the approximation of neglecting the second term in \refeq{lasso-psplit}, which under-estimates \pv{}s. However, we claim the consequent error is small, as justified in theory in \refremark{psplit} and with data in \refsec{results}.
\begin{remark}\textbf{Justification for neglecting the second term in \refeq{lasso-psplit}}:
\label{remark-psplit}
Discovery-aimed variable selection focuses on selecting significant predictor variables, which should have small enough \pv{}s. The second term in \refeq{lasso-psplit} starts at zero and grows as more knots are gone through with decreasing $\lambda$. Significant variables become active earlier, and therefore this approximation incurs much smaller errors on them than in average, especially when the number of genuine predictors is small.
\end{remark}

Remembering the variance of \emph{original} $\vec x_i$ is $\sigma_i^2$, here we define the variances of $\vec x_i$ following the null hypothesis as $\tilde\sigma_i^2$. Propositions are proven in \refsec{proofs}, such as:
\begin{prop}
\eq{\hat\beta_i(\lambda)\ne0\iff\lambda<\tilde\sigma_i\sigma_{\yres}(\lambda)\left|\cor(\vec x_i,\yres(\lambda))\right|.\label{eq-lasso-active}}
\end{prop}
We then consider two possible null hypotheses separately.

\subsection{Normal null hypothesis}
Here, assume the null hypothesis for $\vec x_i=(x_{1,i},x_{2,i},\dots,x_{n,i})^T$ is
\eq{H_0^{(i)}:\ x_{j,i}\sim i.i.d\ N(0,\sigma_i^2),}
The following proposition is proven in \refsec{proofs}:

\begin{prop}For $\vec x\in\bbR^n$ whose elements follow the \textit{i.i.d} standard normal distribution, its dot product squared with another vector $\vec y\in\bbR^n$ follows the $\chi^2$ distribution
\eq{n\sigma_\vec x^2\cor^2(\vec x,\vec y)\sim\chi^2(1),\label{eq-lasso-dist}}
where $\sigma_\vec x^2$ is the variance of $\vec x$.
\end{prop}

Therefore we obtain the \pv\ for every predictor $i$ as
\eqa{&&P\left(\sup\{\lambda:\hat\beta_i(\lambda)\ne0\}\ge\tilde\lambda_i\right)\nonumber\\
&\approx&P\left(\hat\beta_i(\tilde\lambda_i)\ne0\right)\nonumber\\
&=&P\left(\tilde\lambda_i<\tilde\sigma_i\sigma_{\yres}(\tilde\lambda_i)\left|\cor\left(\vec x_i,\yres(\tilde\lambda_i)\right)\right|\right)\nonumber\\
&=&P\left(n\left(\frac{\tilde\sigma_i}{\sigma_i}\right)^2\cor^2\left(\frac{\vec x_i}{\sigma_i},\yres(\tilde\lambda_i)\right)>\frac{n\tilde\lambda_i^2}{\sigma_i^2\sigma_{\yres}^2(\tilde\lambda_i)}\right)\nonumber\\
&=&1-\CDF_{\chi^2(1)}\left(\frac{n\tilde\lambda_i^2}{\sigma_i^2\sigma_{\yres}^2(\tilde\lambda_i)}\right),\label{eq-spv-n}}
where $\CDF_{\chi^2(1)}$ represents the cumulative distribution function for distribution $\chi^2(1)$.

\subsection{Spherical null hypothesis}
Now consider a restrictive null hypothesis, \refeq{h0-s}, which fixes the variance of the null predictor ($\tilde \sigma_i=\sigma_i$). Then we have
\eqa{&&P\left(\sup\{\lambda:\hat\beta_i(\lambda)\ne0\}\ge\tilde\lambda_i\right)\nonumber\\
&\approx&P\left(\hat\beta_i(\tilde\lambda_i)\ne0\right)\nonumber\\
&=&P\left(\left|\cor\left(\vec x_i,\yres(\tilde\lambda_i)\right)\right|>\frac{\tilde\lambda_i}{\sigma_i\sigma_{\yres}(\tilde\lambda_i)}\right)\nonumber\\
&=&2-2\CDF_{t(n-2)}\left(\frac{\tilde\lambda_i}{\sigma_i\sigma_{\yres}(\tilde\lambda_i)}\sqrt{\frac{n-2}{1-\frac{\tilde\lambda_i^2}{\sigma_i^2\sigma_{\yres}^2(\tilde\lambda_i)}}}\right),\label{eq-spv-s}}
where $t(n-2)$ is the Student's $t$-distribution with $n-2$ degrees of freedom.

\begin{remark}Since $\lim_{n\rightarrow\infty}\tilde\sigma_i=\sigma_i$ in the normal null hypothesis, the spherical and normal null hypotheses (and their \pv{}s) converge at large $n$.
\end{remark}

\begin{remark}The \pv\ of the first predictor under the spherical null hypothesis is the same with that of its Pearson correlation with the target vector $\vec y$.
\end{remark}

\section{Proofs}
\label{sec-proofs}
\subsection{Proposition 1}
First prove
\eq{\hat\beta_i(\lambda)>0\iff\lambda<\tilde\sigma_i\sigma_{\yres}(\lambda)\cor(\vec x_i,\yres(\lambda)).}
Define
\eq{\tilde\beta_j(\lambda)\equiv\left\{\begin{array}{ll}\alpha,&\mathrm{for\ }j=i,\\\hat\beta_j(\lambda),&\mathrm{else}.\end{array}\right.}
Since
\eq{\hat \beta_i(\lambda)\equiv\argmin_\alpha\frac{1}{2n}||\vec y-\vec X\tilde\beta||_2^2+\lambda||\tilde\beta||_1,}
then
\eqa{&&\hat \beta_i(\lambda)>0\nonumber\\
&\iff&\left.\frac{\partial}{\partial\alpha}\left(\frac{1}{2n}||\vec y-\vec X\tilde\beta(\lambda)||_2^2+\lambda||\tilde\beta(\lambda)||_1\right)\right|_{\alpha=0^+}<0\nonumber\\
&\iff&\left.\frac{1}{n}\vec x_i^T(\alpha\vec x_i-\yres(\lambda))+\lambda\right|_{\alpha=0^+}<0\nonumber\\
&\iff&\lambda<\tilde\sigma_i\sigma_{\yres}(\lambda)\cor(\vec x_i,\yres(\lambda)).}
Similarly,
\eq{\hat\beta_i(\lambda)<0\iff\lambda<-\tilde\sigma_i\sigma_{\yres}(\lambda)\cor(\vec x_i,\yres(\lambda)).}
Since $\hat\beta_i(\lambda)$ and $\cor(\vec x_i,\yres(\lambda))$ always have the same sign, we can combine the two as
\eq{\hat\beta_i(\lambda)\ne0\iff\lambda<\tilde\sigma_i\sigma_{\yres}(\lambda)\left|\cor(\vec x_i,\yres(\lambda))\right|.}

\subsection{Proposition 2}
Due to the rotational $SO(n)$ symmetry in the PDF of $\vec x\in \bbR^n$, each of which follows i.i.d $N(0,1)$, the distribution of $n\sigma_\vec x^2\cor^2(\vec x,\vec y)$ does not depend on $\vec y$. For simplicity, choose
\eq{y_1=\sqrt{n/2},\hspace{2em}y_2=-\sqrt{n/2},\hspace{2em}y_i=0,\mathrm{\ for\ }i=3,\dots,n.}
Then, by expanding the correlation, we have
\eq{n\sigma_\vec x^2\cor^2(\vec x,\vec y)=(x_1/\sqrt2-x_2/\sqrt2)^2.}
Since $x_1,x_2\sim i.i.d\ N(0,1)$, define
\eq{z\equiv x_1/\sqrt2-x_2/\sqrt2\sim N(0,1).}
Therefore
\eq{n\sigma_\vec x^2\cor^2(\vec x,\vec y)=z^2\sim\chi^2(1).}

\end{document}